\documentclass[]{aastex63}

\newcommand{\comment}[1]{}

\shorttitle{Deuterium Escape on Sub-Neptunes }
\shortauthors{Gu \& Chen}

\begin{document}

\title{Deuterium Escape on Photoevaporating Sub-Neptunes}

\correspondingauthor{Pin-Gao Gu}
\email{gu@asiaa.sinica.edu.tw}

\author{Pin-Gao Gu}
\affiliation{Institute of Astronomy \& Astrophysics, Academia Sinica,
Taipei 10617, Taiwan}

\author{Howard Chen}
\affiliation{Department of Aerospace, Physics, and Space Sciences, Florida Institute of Technology, Melbourne, FL 32901, USA}
\affiliation{Planetary Environments Laboratory, NASA Goddard Space Flight Center,  Greenbelt, MD 20771, USA}







\begin{abstract}
We investigate the evolution of the deuterium-to-hydrogen (D/H) mass ratio driven by EUV photoevaporation of hydrogen-rich atmospheres of close-in sub-Neptunes around solar-type stars. 
For the first time, the diffusion-limited approach in conjunction with energy-limited photoevaporation is considered in evaluating deuterium escape from evolving exoplanet H/He envelopes.
We find that 
the planets with smaller initial gas envelopes and thus smaller sizes can lead to weaker atmospheric escape, which facilitates hydrogen-deuterium fractionation.
Specifically, in our grid of simulations with low envelope mass fraction less than 0.005, a low-mass sub-Neptune (4-$5M_\oplus$) at about 0.25-0.4 au or a high-mass sub-Neptune (10-$15M_\oplus$) at about 0.1-0.25 au can increase the D/H values by greater than 20\%  over 7.5 Gyr.
Akin to the helium-enhanced envelopes of sub-Neptunes due to photoevaporating escape, the planets along the upper boundary of the radius valley are the best targets to detect high D/H ratios. 
The ratio can rise by a factor of $\lesssim$ 1.65 within 7.5 Gyrs in our grid of evolutionary calculations. The
D/H ratio is expected to be higher in thinner envelopes as long as the planets do not become bare rocky cores.
\end{abstract}



\section{Introduction} 
\label{sec:intro}
Deuterium has long been one of the most studied isotopes in terms of its abundance relative to hydrogen in a variety of astronomical environments, despite being a trace element in the Universe. 
Deuterium 
was largely produced in the Big Bang. The primordial deuterium-to-hydrogen (D/H) ratio inferred from observations of the high-redshift intergalactic medium and cosmic microwave background radiation constrained the baryon abundance in the Big Bang nucleosynthesis and cosmological parameters \citep[e.g.,][]{Tytler96,Planck}. 
The D/H ratio 
in the present local interstellar medium was inferred through Lyman-$\alpha$ line observations \citep[e.g.,][]{Linsky95}, which are lower than the protosolar value 4.55 Gyr ago implied from solar wind measurements \citep[e.g.,][]{GG98}. This decrease in D/H over time is expected because deuterium has been destroyed by nucleosynthesis in stars \citep[e.g.,][]{Lellouch2001}. By contrast,
the D/H values can be considerably high in prestellar cores and protoplanetary disks where the temperature is low enough to form deuterated molecules, 
which in turn enrich the icy dust mantle
through grain-surface chemistry \citep[e.g.,][]{Cleeves14,Ceccarelli14}. Furthermore, within the Solar System, the D/H ratios were measured or observed in chondrites,  comets, planetary atmospheres, and Earth's ocean water to investigate how the different ratios are possibly evolved through various fractionation processes \citep[e.g.,][]{Morley19,Atreya20,Piani2020,Nomura23}.

In the context of formation of giant planets in the Solar System, the D/H ratio of Jupiter and Saturn is comparable to the  protosolar value $\approx 2\times 10^{-5}$ \citep[e.g.,][]{GG98,Pierel17}, roughly consistent with formation through massive gas accretion from the solar nebula. On the other hand, the D/H ratio of Uranus and Neptune is higher than the protosolar value by a factor of $\sim 2.5$ \citep{Feucht13}. The theoretical interpretation is unclear \citep[e.g.,][]{GG14}. It could be a natural consequence of icy giant planets, which accrete 
the nebular gas and a significant amount of D‐enriched ice during their formation
\citep[e.g.,][]{Watson74,GR81,Lecluse96,Drouart99,Hersant01}.

For terrestrial planets, atmospheric escape has been posited to explain their current D/H ratios. The D/H ratio in Venus's atmosphere is much higher than that of Earth's ocean ($\approx 1.56\times 10^{-4}$) by about a factor of 100. Previous work has shown that the rapid evaporation of the putative early ocean via runaway greenhouse effect 
can possibly induce more loss of hydrogen to space than the heavier isotope D, leading to the high D/H ratio on Venus \citep{DP83,Kulikov06}. The scenario of the runaway greenhouse effect has been theorized to delineate the inner boundary of the circumstellar habitable zone, where liquid water can exist on the crust of a temperate planet in a range of orbital distances \citep{Kasting93,Kopp13,Kasting15}. In the cases of Earth and Mars, their current D/H ratios are similar to those of carbonaceous chondrites and Oort cloud comets, respectively. Assuming that these ratios have remained constant for billions of years, the ratios can be attributed to the late accretion of the carbonaceous asteroids and comets that were delivered from the outer Solar System at or near the end of the stage of planet formation \citep[e.g.,][]{MA86,Gomes05,Drake05,RI17}. However, atmospheric escape could further sculpt the D/H ratio after the completion of planet formation through the accretion.
Analogous to Venus, early Mars may also have surface water as the major reservoir of hydrogen \citep[e.g.,][]{Ramirez14,Batalha15}. As a result, the loss of water and atmosphere from Mars' surface could bring the D/H ratio to 5--7 times larger than the terrestrial value \citep{Kras15,Villa15}. Based on formation modeling or geochemical evidence, the early Earth could gravitationally accrete hydrogen gas from the remanent solar nebula to form its primordial atmosphere \citep{IG06,Marty12,LC16}. The subsequent loss of this H-rich atmosphere would raise the protosolar D/H ratio to the current value, followed by deuterium exchange between hydrogen gas and water vapor during the ocean formation \citep{GI08}. The ingassing process of the H-rich atmosphere to Earth's interior was also posited to contribute to some of the water content in Earth's core during the impact phase of growing planetary embryos \citep{Wu18}\footnote{An early steam atmosphere could also have been lost via impact erosion by planetesimals during planetary accretion (e.g., \citealt{CJ22}), which could lead to degassing of mantle hydrogen and removal from the atmospheric reservoir.}. 

Beyond the solar system, atmospheric escape plays a major role in the demographic, evolutionary, and observational narratives of exoplanets.
The atmospheric loss from close-in exoplanets of Neptune to Jupiter masses has been observed during planet transits \citep[e.g.,][]{Vidal03,Ehren15,Spake18}. The transit depths indicated by the broad profiles of Lyman-$\alpha$ and He triple lines are greatly enhanced, implying that the H-rich atmosphere can reach the Roche lobe with a velocity beyond the escape velocity of these close-in planets \citep[e.g.,][for a recent review]{Owen19}. Additionally, the {\it Kepler} transit survey \citep{Borucki10} in conjunction  with follow-up spectroscopic observations and parallax measurements has revealed the so-called ``radius valley", i.e., the low occurrence rate for super-Earths of planetary radii $\sim 1.8 R_\oplus$ within the orbital period of $\sim 100$ days around FGK dwarfs \citep{Fulton17,FP18}. The radius valley was also identified by asteroseismic measurements for a sample of bright planet-hosting stars, of which the stellar radii were more accurately determined \citep{Van18}. The bimodal distribution of the planet radius separated by this radius valley is consistent with the photoevaporation and core-powered mass-loss models \citep[e.g.,][]{OW13,Jin2014,OW17,Ginzburg18,GS19}. In these atmosphere loss models, 
a sub-Neptune consists of a hydrogen-rich envelope and a rocky core. When the envelope mass fraction is about a few percent of the total planet mass, the mass-loss timescale becomes large enough for a sub-Neptune to survive
\citep[e.g.,][]{CR16,OW17,Ginzburg18}.\footnote{ More specifically, the mass-loss timescale needs to be larger than the radiative cooling timescale for a contracting envelope in the core-powered mass-loss model \citep{Ginzburg18,GS19}.}
Therefore,
the rocky planets larger than the radius valley can retain the H-rich atmosphere against atmospheric escape over billions of years, whereas the rocky planets smaller than the radius valley, i.e., super-Earths, almost lose their primordial atmosphere and become almost bare.

Alternatively, the {\it Kepler} small planets larger than the radius valley could be massive icy planets with a thin H atmosphere, which could have been partially sculpted by atmospheric escape as well during planetary evolution \citep{Zeng19,Ven20}. Given the composition degeneracy, \citet{MLM21} subsequently modeled a wide range of interior compositions and envelope mass fractions of extreme-ultraviolet (EUV) photoevaporative planets compared to the observed bimodal size distribution. The authors reached a similar conclusion to previous models for the compositions of super-Earths and sub-Neptunes, with a more stringent constraint on the initial mass distribution of these planets.
Furthermore, the atmospheric photoevaporation was applied to the post-evolution of the final assembled planets through giant impacts in the attempt to simultaneously model the similarity of planet size and orbital spacing in multiple planetary systems (a.k.a. peas-in-a-pod, see a review by \citet{Weiss23}), as well as the radius valley for the {\it Kepler} small planets \citep{matsumoto21}. 

As the loss of the H-rich atmosphere of terrestrial planets can produce element fractionation in the Solar System, a similar effect is expected to happen in exoplanets. To explain the lack of CH$_4$ and dominance of CO in the atmosphere of GJ 436b, \citet{Hu15} proposed that the atmospheric loss from warm Neptune- and sub-Neptune-sized exoplanets can strongly induce diffusive separation of hydrogen and helium, leading to a preferential escape of hydrogen and thereby yielding a helium-dominated atmosphere.  
Based on the extension to the MESA\footnote{The Modules for Experiments in Stellar Astrophysics \citep{Paxton11,Paxton13,Paxton15,Paxton18,Paxton19}; also refer to
\url{https://docs.mesastar.org}.} module that \citet{CR16} initially developed,
\citet{MR20} revisited the problem by 
considering the coevolution of the envelope photoevaporation and planet radius. The compositional simulation coupled with thermal structure enabled the authors to show that GJ 436b is too large to possess a He-rich atmosphere around a rocky core. Nevertheless,  the fractionation effect proposed by \citet{Hu15} can still be possible in the atmosphere of highly irradiated sub-Neptunes with low envelope fractions.  The predictive consequence is that the atmospheres of sub-Neptunes along the edge of the radius valley may be helium-enhanced \citep{MR23}. If this is the case, however, then other light elements and isotopes could also be subjected to dramatic modulations by atmospheric loss.

Motivated by the immense interest in the H-D fractionation
for the solar-system planets (i.e., formation, atmospheric evolution, and water delivery),
potential observations of deuterium fractionation for exoplanets have also been posited. This would be achieved by detecting
isotopic molecules in exoplanet atmospheres 
with ongoing and upcoming facilities, such as the James Webb Space Telescope ({\it JWST}) and the Extremely Large Telescope \citep{KV19,Lincowski19,MS19,Morley19}. Despite these recent observational efforts and propositions, there is as yet no theoretical study to model H-D fractionation processes for exoplanets  with significant H/He envelopes. In this Letter, we 
     examine the evolution of the D/H ratio for close-in sub-Neptunes due to atmospheric escape based upon the aforementioned 
modeling for the H-He fractionation driven by the photoevaporation of a primordial H-rich envelope. 

\section{Fractionation equations including deuterium escape}

Using MESA version 12778, we 
simulate the coupled thermal and compositional evolution of photoevaporating sub-Neptunes. The code we employed has been  substantially modified based on the module set up by \citet{MR23}; i.e.,  we add the diffusive escape of deuterium in addition to the  H-He fractionation scheme.
We consider a fiducial model for a solar-type star of surface temperature 6000 K, a planet Bond albedo equal to 0.2, and the same initial entropy of the planetary gas envelope as that specified by \citet{MR20}. 
\citet{MR20} found that the enhancement of helium fraction due to the preferential hydrogen loss can shape the mass-radius relation of the sub-Neptune-mass planet population. The effect is more prominent for low-mass ($M_p \lesssim 10 M_\oplus$) highly irradiated planets with an initial envelope mass fraction $f_{env}$ below 1.0\% (see Figure 3 of \citet{MR20}). Because deuterium's mass lies between the masses of H and He,  it is expected that H and D fractionation due to photoevaporation can also occur on close-in sub-Neptunes.

We begin the calculation of deuterium escape with the photoevaporative wind described by Equation(\ref{eq:orig}) in Appendix \ref{sec:app1}.
After the inclusion of D in Equation(\ref{eq:orig}, Equations (\ref{eq:Phi}) and (\ref{eq:He}) become\footnote{It can be argued that the energy $f_r \Phi_{EL}$ is deposited differently among H, D, and He.}
\begin{eqnarray}
&& \Phi \approx \Phi_H + \Phi_D + \Phi_{He}=4 \pi R_h^2 (\phi_H m_H + \phi_D m_D +\phi_{He} m_{He}), \label{eq:cont}\\
&& {\phi_{He} \over X_{He}} \approx {\phi_H \over X_H} - \phi_{DL,He} + \left( {1\over b_{D,He}} - {1\over b_{D,H}} \right) b'_{H,He} \phi_D, \label{eq:He_D}
\end{eqnarray}
where the escape rate $\Phi$ is given by Equation(\ref{eq:Phi_EUV}) in Appendix \ref{sec:app2}, 
$\phi$ is the escaping number flux evaluated at the homopause radius $R_h$,  $m$ is the atomic mass of a species, $X$ is the mixing ratio of a species, $b$ is the binary diffusion coefficient between two species, $b'_{H,He}$ is $b_{H,He}$ taking into
account the H ionization, and $\phi_{DL,He}$ is the diffusion-limited escape flux for He. Here $R_h$ is determined by the location where the eddy diffusivity $K_{zz}$ equals the binary diffusivity $\mathcal{D}_{H,He}$. Detailed descriptions of these quantities can be found in Appendix \ref{sec:app2}.
In deriving Equation(\ref{eq:He_D}), we have assumed $X_D \ll (b_{D,He}/b_{He,H}) X_H$, $X_D \ll (b_{D,H}/b_{He,H}) X_H$, and $X_H+X_{He} \approx 1$. Because $1/b_{D,He}-1/b_{D,H} >1$ (see below), the outflow of D may reduce the mass fractionation between H and He due solely to the gravity (i.e., diffusion-limited) effect given by $\phi_{DL,He}$. Furthermore, in Equation({\ref{eq:He_D}), the term due to $\phi_D$ is likely much smaller than $\phi_H/X_H$ and could be ignored. In this case, Equation({\ref{eq:He_D}) can just be approximated to Equation(\ref{eq:He}).

Similarly, the number flux of D from Equation(\ref{eq:orig}) is given by
\begin{equation}
{\phi_D \over X_D}\approx {\phi_H-{GM_p(m_D-m_H) \over kT r_0^2}b_{D,H}  + {b_{D,H} \over b_{He,H}}\left( {X_{He}\over X_H}-{\phi_{He}\over \phi_H} \right)\phi_H + {b_{D,H}\over b_{D,He}}\phi_{He} \over X_H+{b_{D,H} \over b_{D,He}}X_{He}}, \label{eq1:D}
\end{equation}
where  $G$ is the gravitational constant, $M_p$ is the planetary mass, $T$ is the temperature, $k$ is the Boltzmann constant, $r_0$ is the homopause radius for evaluating $\phi_H$, $\phi_D$, and $\phi_{He}$, and $X_D \ll X_H$ has been applied to simplify the equation.
If $\phi_{He} \neq 0$, the above equation along with Equation(\ref{eq:He_D}) can be rewritten as
 \begin{equation}
 {\phi_D \over X_D} \approx  {\phi_H-\phi_{DL,D}  + \alpha_2 \phi_{DL,He}X_{He} + \alpha_3 \phi_{He} \over X_H+\alpha_3 X_{He}},\label{eq2:D}
\end{equation}
where
\begin{equation}
\phi_{DL,D} \equiv {GM_p (m_D-m_H) b'_{D,H} \over kT r_0^2},\qquad
\alpha_2 \equiv {b'_{D,H} \over b'_{He,H}}, \qquad
\alpha_3 \equiv {b'_{D,H} \over b_{D,He}}.  \label{eq2_1:D}
\end{equation}
Note that in the above expressions, some of the $b$ coefficients have been replaced with $b'$ to take into account the H ionization (see below). The four terms on the right-hand side of Equation(\ref{eq2:D}) can be physically interpreted as hydrogen drag (first term), helium drag (fourth term), and the diffusion relative to hydrogen due to gravity (second term), which is mitigated by the diffusion between hydrogen and helium due to gravity (third term).
It is evident from Equation(\ref{eq2:D}) that $\phi_D/X_D \rightarrow \phi_H/X_H$ as $b'_{D,H} \rightarrow 0$ or as both $\phi_{DL,D}$ and $\phi_{DL,He} \rightarrow 0$ (i.e., no H-D fractionation when D and H are strongly coupled). In addition,  when $X_{He}=0$ and thus $X_H \approx 1$, Equation(\ref{eq2:D}) becomes the same as Equation(\ref{eq:He}) with He replaced by D, as expected. 
For simplicity, we assume $r_0$ to be the homosphere for hydrogen and helium $R_h$; hence, $T$ equals the homopause temperature $T_h$. This assumption will be discussed near the end of the paper.

Therefore, Equations(\ref{eq:cont}), (\ref{eq:He_D}), and (\ref{eq1:D}) (or eq(\ref{eq2:D}) if $\Phi_{He}$ is nonzero) can be solved for $\phi_H$, $\phi_{He}$, and $\phi_{D}$ evolving with time as the envelope escapes at the rate given by $\Phi$. Since deuterium is a trace element with $X_D \sim 10^{-5}$ to $10^{-4}$, it is expected that $\Phi_H$ and $\Phi_{He}$ would not change noticeably due to the presence of D. Hence, it is  an excellent approximation to use the same expressions for H and He mass loss as those derived by \citet{Hu15}. The mass-loss rates of H, He, and D then read

If $\Phi \leq \Phi_{crit,He} \equiv \phi_{DL,He} X_H m_H 4 \pi R_h^2$,
\begin{eqnarray}
\Phi_H &\approx& \Phi,\label{eq:He_crit1}\\
\Phi_{He}& \approx &0,\\
\Phi_D&\approx &X_D m_D { (1+ \alpha_2 X_{He}/X_H) \Phi_H/m_H - \phi_{DL,D}4\pi R_h^2 \over X_H + \alpha_3 X_{He}}, \label{eq:Phi_D_noHe}\\
&&{\rm however\ if\ } \Phi \leq \Phi_{crit,D} \equiv {1 \over 1+ \alpha_2 X_{He}/X_H}\phi_{DL,D} 4\pi R_h^2 m_H ,\ \Phi_D=0. 
\label{eq:He_crit4}
\end{eqnarray}

If $\Phi > \Phi_{crit,He}$
\begin{eqnarray}
\Phi_H&\approx &{\Phi m_H X_H + \phi_{DL,He} m_H m_{He} X_H X_{He} 4 \pi R_h^2 \over m_H X_H + m_{He} X_{He}},\\
\Phi_{He}&\approx &{\Phi m_{He} X_{He} - \phi_{DL,He} m_H m_{He} X_H X_{He} 4 \pi R_h^2 \over m_H X_H + m_{He} X_{He}},\label{eq:Phi_He}\\
\Phi_D &\approx & X_D m_D {\Phi_H/m_H - \phi_{DL,D}4\pi R_h^2 + \alpha_2 \phi_{DL,He} X_{He} 4 \pi R_h^2 +\alpha_3 \Phi_{He}/m_{He} \over X_H + \alpha_3 X_{He}}. \label{eq:Phi_D}
\end{eqnarray}
Note that if we ignore the He species (i.e., $X_{He}=\Phi_{He}=0$), Equations(\ref{eq:Phi_D_noHe}) and (\ref{eq:Phi_D}) are identical and have the same form as Equation(\ref{eq:Phi_He}) with He replaced by D. In this case, without He, $\Phi_D=0$ when $\Phi \leq \phi_{DL,D} 4 \pi R_h^2 m_H$, analogous to the criterion for the He escape with He replaced by D and $X_H \approx 1$.

We adopt the following values for diffusion coefficients: 
$b_{H,He}=1.04\times 10^{18} T_h^{0.732}$ cm$^{-1}$ s$^{-1}$  \citep{Hu15} and $b_{D,H}=7.183\times 10^{17} (T_h)^{0.728}$ cm$^{-1}$ s$^{-1}$ \citep{GI08}. $b_{D,He}$ is crudely estimated from $b_{D,H}$ using the square root of the reduced mass \citep{KP83,GI08}, i.e., $b_{D,He}/b_{D,H} \approx  \sqrt{\left( {m_D+m_{He} \over m_D m_{He}} \right)  /   
\left( {m_D+m_{H} \over m_D m_{H}} \right) } \approx$ 0.7,
which validates the aforementioned relation  $1/b_{D,He}-1/b_{D,H} >1$ in Equation(\ref{eq:He_D}).
The correction from ionized species can be made following \citet{Hu15}. Specifically,
the binary diffusion coefficient $b_{D,H}$ corrected for H ionization, denoted by $b'_{D,H}$, is given by the relation
$kT /b'_{D,H}=(1-x)kT /b_{D,H}+x m_{H^+} \nu_{D,H^+} /n_D$,
where $x$ is the ionization fraction of H and the momentum transfer collisional frequency $ \nu_{i,H^+}  = n_i \langle \sigma v \rangle_{i,H^+} m_i/(m_i+m_{H^+})$ for the neutral species denoted by $i$. 
We use $\nu_{He,H^+}/n_{He}=10.6\times 10^{-10}$ cm$^3$/s \citep{SN80,Hu15} to estimate $\nu_{D,H^+} \approx \sqrt{m_{H^+}m_D/(m_{H^+}+m_D) \over m_{H^+}m_{He}/(m_{H^+}+m_{He})}\nu_{He,H^+} (n_D/n_{He})$. 
In the MESA module by \citet{MR23}, $x$ is determined by the Saha equation ignoring photoionization. 
These values are expected to be insensitive to the H ionization fraction $x$ \citep{Hu15}. In this study, we ignore deuterium ionization and its effect on the binary diffusion coefficients for simplicity, as its momentum transfer collision frequencies with hydrogen and helium are unknown. 

Following a similar procedure for evolving $X_H$, $X_{He}$ and $X_{Z_i}$ by \citet{MR20}, we can evolve the abundance of D as follows:
 \begin{equation}
 X_{D,n}={M_{env,n-1} X_{D,n-1} - (\Phi_D dt) \over M_{env,n-1}-(\Phi_{He}+\Phi_H+\Phi_D)dt},
 \label{eq:X_D}
 \end{equation}
 where $n$ is the index for the time step. We consider the same elemental species  as those in \citet{MR20}\footnote{Eight elemental species are considered in \citet{MR20}: $^1$H, $^3$He, $^4$He, $^{12}$C, $^{14}$N, $^{16}$O, $^{20}$Ne, and $^{24}$Mg.} and add deuterium to the reaction network of MESA.
The $^3$He escape is not modeled because its effect on the deuterium escape is expected to be negligible due to its extremely low abundance compared to H and He.






\section{Results}

We find that stellar EUV-induced hydrodynamic escape can elevate H-D fractionation in the hydrogen-dominated envelopes of our modeled {\it Kepler} planets; in many scenarios, the elevation in D/H is greater than 20\% over 7.5 Gyr.
 Unless otherwise stated, 
we follow \citet{MR20} and adopt the fiducial values for the free parameters for the EUV photoevaporation efficiency ($\eta=0.1$) and homopause conditions ($K_{zz}=10^9$ cm$^2$/s and $T_h=10000$ K) in this study. 
Changing the values of these free parameters has a minor influence on our results for D/H (see Appendix \ref{sec:app3}).

\subsection{Typical evolution}

\begin{figure}
\centering
\setlength{\tabcolsep}{0pt}
\begin{tabular}{ccc}
\includegraphics[width=0.35\linewidth]{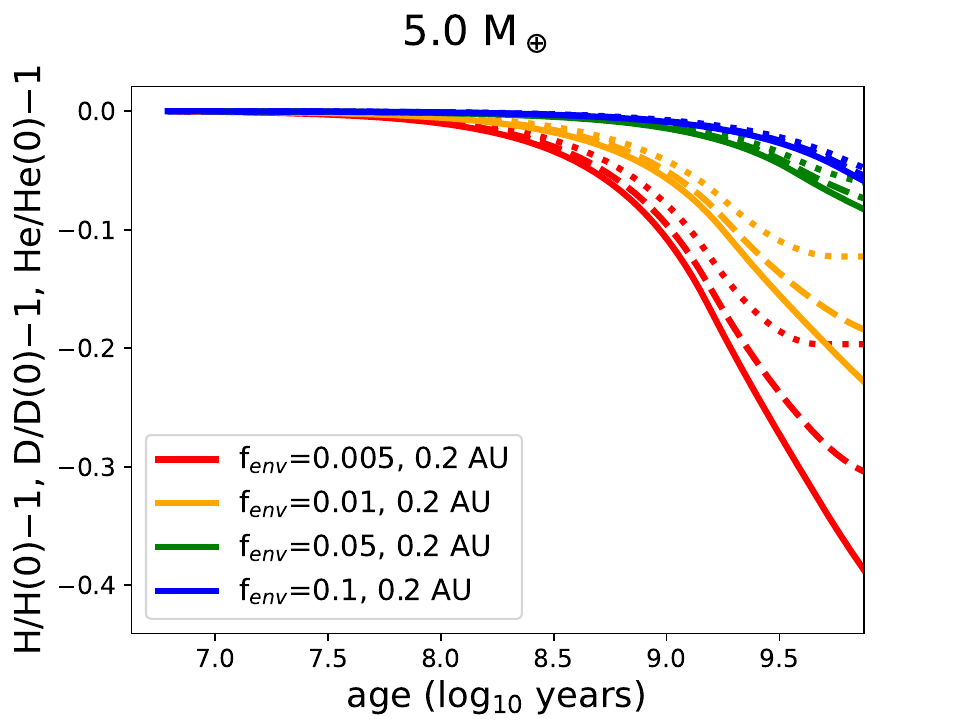} &
\includegraphics[width=0.35\linewidth]{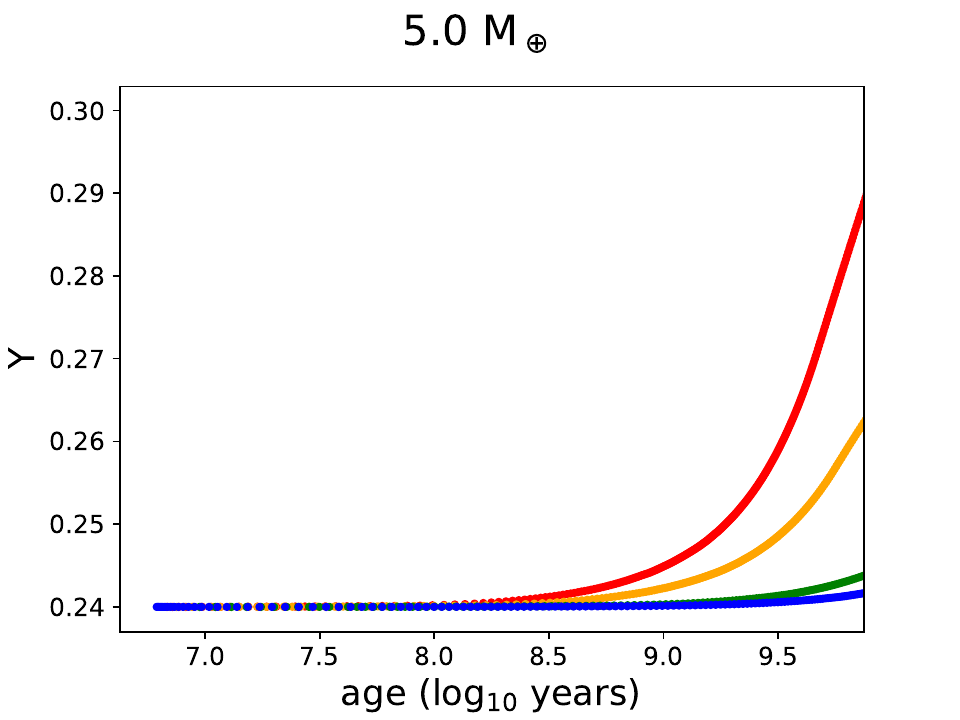}  &
\includegraphics[width=0.35\linewidth]{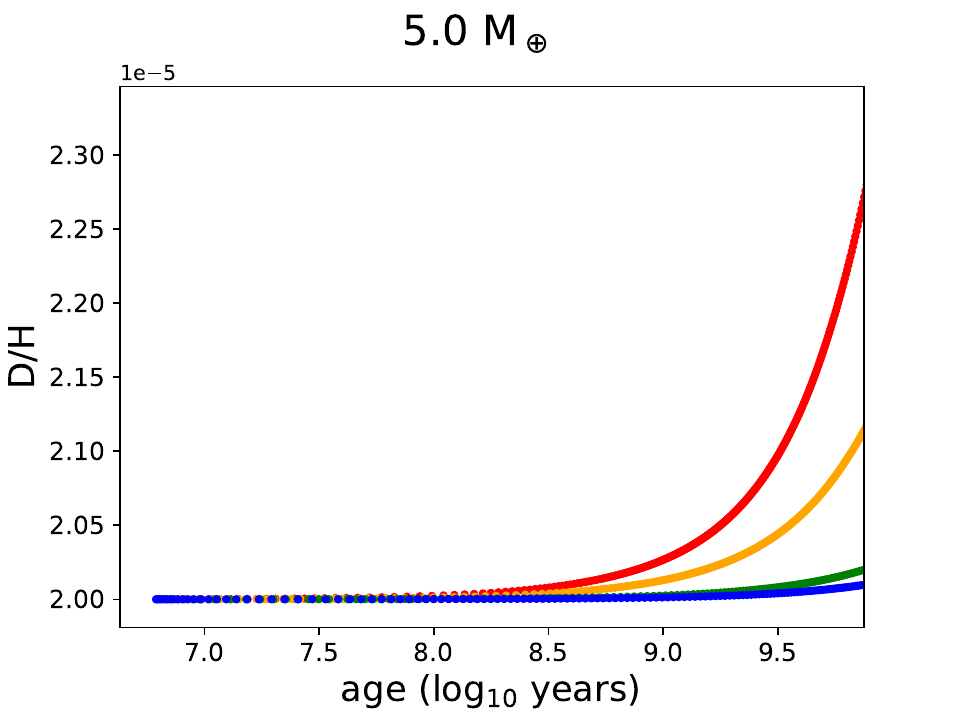} \\
\includegraphics[width=0.35\linewidth]{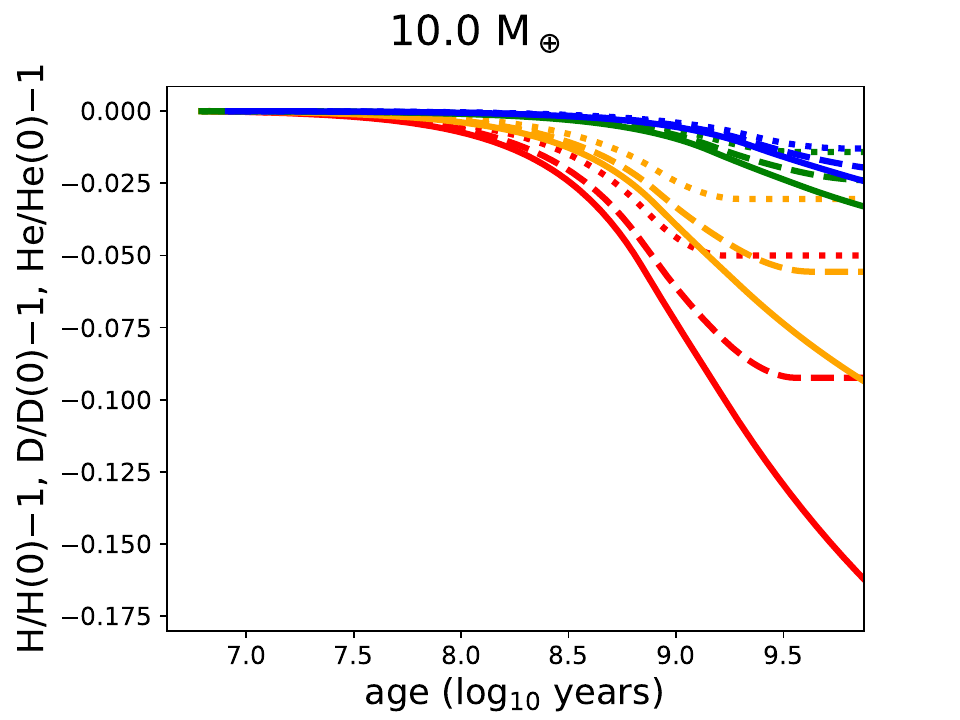} &
\includegraphics[width=0.35\linewidth]{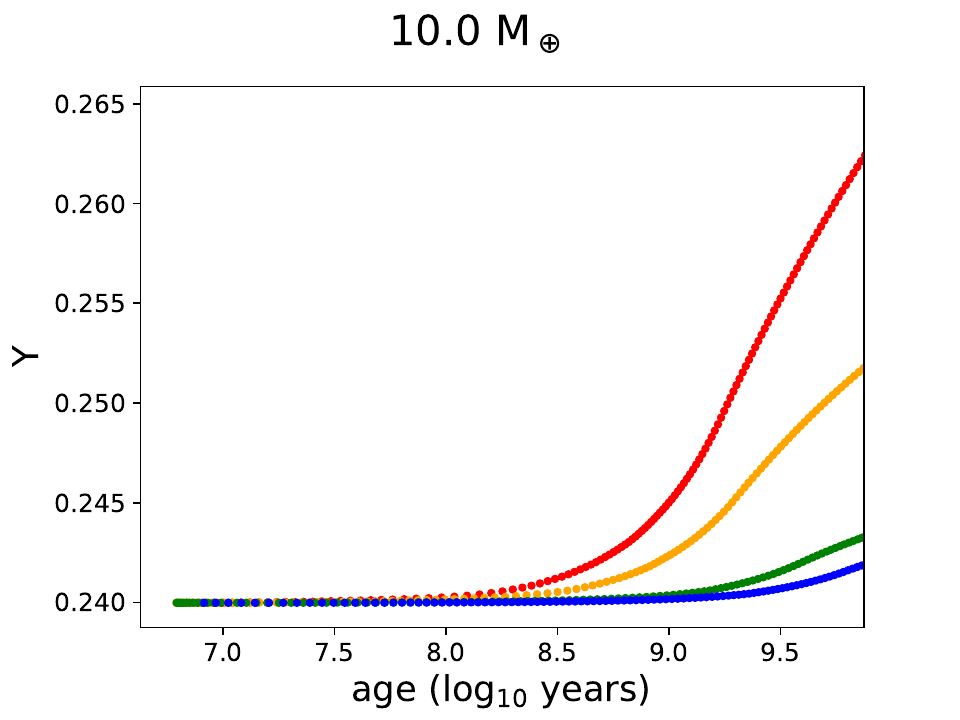}  &
\includegraphics[width=0.35\linewidth]{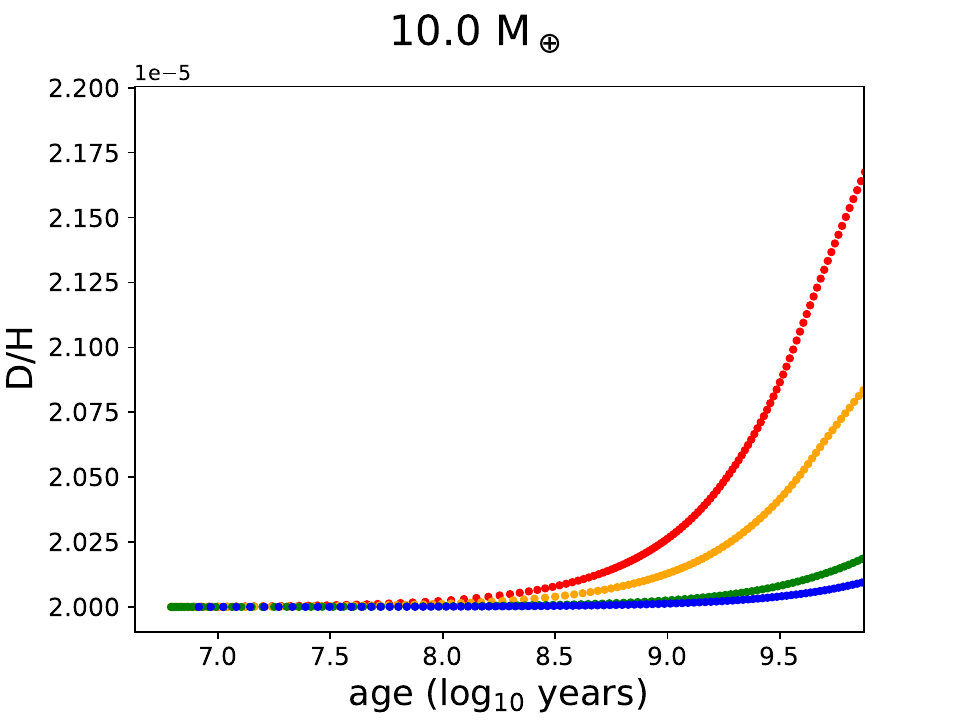}
\end{tabular}
\caption{Fractionation evolution of the envelope for 5 and 10 $M_\oplus$ at $d=0.2$ au with various initial mass fractions of the envelope $f_{evn}$. Left panels: mass-loss evolution of hydrogen ($X_H/X_{H,0}-1$, solid), deuterium ($X_{D}/X_{D,0}-1$, dashed), and helium ($X_{He}/X_{He,0}-1$, dotted). Middle panels: Y evolution. Right panels: D/H evolution. The fractionation between H, D, and He becomes more  substantial at later times due to the decrease in the mass-loss rate, leading to the increases in Y and D/H with time. The trend is more prominent for the planet with a smaller initial mass fraction of the gas envelope $f_{env}$ due to the even lower mass-loss rate. Notably, the mass loss of He or D can almost stop at later times, as indicated by the flat curves in the left panels.}
\label{fig:evol_0.2}
\end{figure}

The inclusion of deuterium escape into the compositional evolution of super-Earths and sub-Neptunes allows us to simultaneously track the variation of hydrogen, helium, and deuterium with time.
Example evolution tracks of a 
5 and 10 $M_\oplus$ planet with different $f_{env}$ at $d=0.2$ au experiencing photoevaporation-driven fractionation are shown in Figure~\ref{fig:evol_0.2}. The left panels plot the deviation from the initial abundances, the middle panels present the helium abundance Y (i.e., $X_{He}$), and the right panels display the D/H ratio.  
As illustrated in the left panels of Figure~\ref{fig:evol_0.2},  while the helium loss fraction $X_{He}/X_{He,0}-1$ is clearly less than the hydrogen loss fraction $X_H/X_{H,0}-1$  \citep{MR20,MR23}, the deuterium loss fraction $X_D/X_{D,0}-1$ is also fairly distinguishable; it is less than the hydrogen but more than the helium loss fraction, as expected from their mass differences.
The abundance deviations of H, D, and He from their initial values become more  substantial at later times. This change is expected. As the gas envelope contracts over time and the EUV flux from the star also decreases substantially after 1 Gyr, 
the mass-loss rate $\Phi$ decreases with time according to Equation(\ref{eq:Phi_EUV}). 

Planets with lower $f_{env}$ experience the most pronounced change in their H, D, and He abundance compared to their initial values.
A planet with lower $f_{env}$ has smaller cross-sectional radius and thus a lower 
mass-loss rate during the evolution.
Notably, the flat curves for $f_{env}=0.005$ and 0.01 at the late stage in the bottom left panel of Figure~\ref{fig:evol_0.2} indicate that at the age of $\sim $ a few Gyrs, the escape of deuterium and helium halts with the escape of H in the atmosphere of the planet with 10 $M_\oplus$. This is because the total mass-loss rate $\Phi$ starts to become smaller than both the critical values for the helium loss rate $\Phi_{crit,He}$ and the deuterium loss rate $\Phi_{crit,D}$ at this point, as described by Equations~(\ref{eq:He_crit1})-(\ref{eq:He_crit4}). In comparison, $\Phi$ has not reached its critical value for deuterium in the case of 5 $M_\oplus$ within 7.5 Gyr, primarily due to the higher mass-loss rate from the weaker gravitational potential of a less massive planet. 

As shown in the left panels of Figure~\ref{fig:evol_0.2},
hydrogen 
can be lost by more (less) than 20\% for $f_{evn} \leq 0.01$ of a planet with 5 $M_\oplus$ (10 $M_\oplus$), and deuterium
can be lost by more (less) than 10\% for $f_{evn} \leq 0.01$ of a planet with 5 $M_\oplus$ (10 $M_\oplus$).  Consequently,  for a planet of  5 (10) $M_\oplus$ with $f_{evn}=0.005$-0.01,
the helium abundance Y increases to $\approx 0.26$-0.29 (0.25-0.26) from the initial value 0.24 and the D/H ratio increases to  $\approx 2.12$-2.3 (2.08-2.17) $\times 10^{-5}$ from the initial ratio 2$\times 10^{-5}$ within 7.5 Gyr, as illustrated in the middle and right panels of  Figure~\ref{fig:evol_0.2}. Although all of our models start from $2\times 10^{-5}$ for the D/H ratio of the protosolar value, it is worth mentioning that the deuterium abundance $X_D$ should almost scale as its initial value
for a given planetary mass and $f_{env}$. This is because deuterium is a trace element; thus, 
its variability hardly changes
the gas envelope mass during evolution. This can also be realized from Equation(\ref{eq:X_D}) by noticing $\Phi_D \ll \Phi_{H}+\Phi_{He}$ and $\Phi_D \propto X_{D,n-1}$. Therefore, for a planet of 5 (10) $M_\oplus$ with $f_{env}=0.005$-0.01, the D/H ratios increase by a factor of $\approx$ 6-15\% (4-8.5\%) within 7.5 Gyr, regardless of the initial deuterium abundance.


\begin{figure}
\centering
\setlength{\tabcolsep}{0pt}
\begin{tabular}{ccc}
\includegraphics[width=0.35\linewidth]{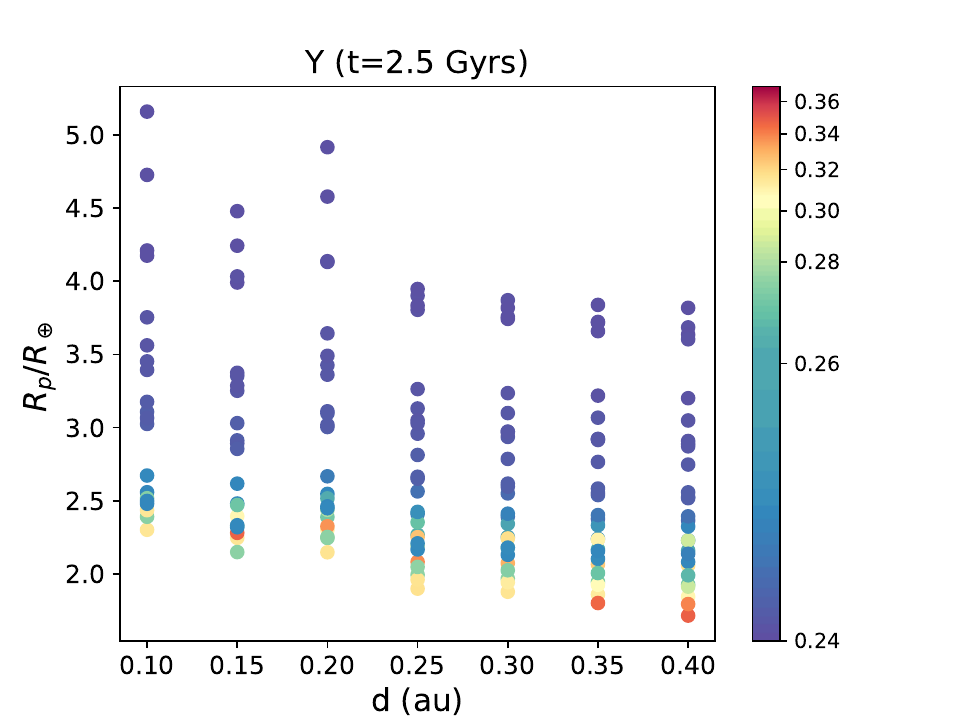} &
\includegraphics[width=0.35\linewidth]{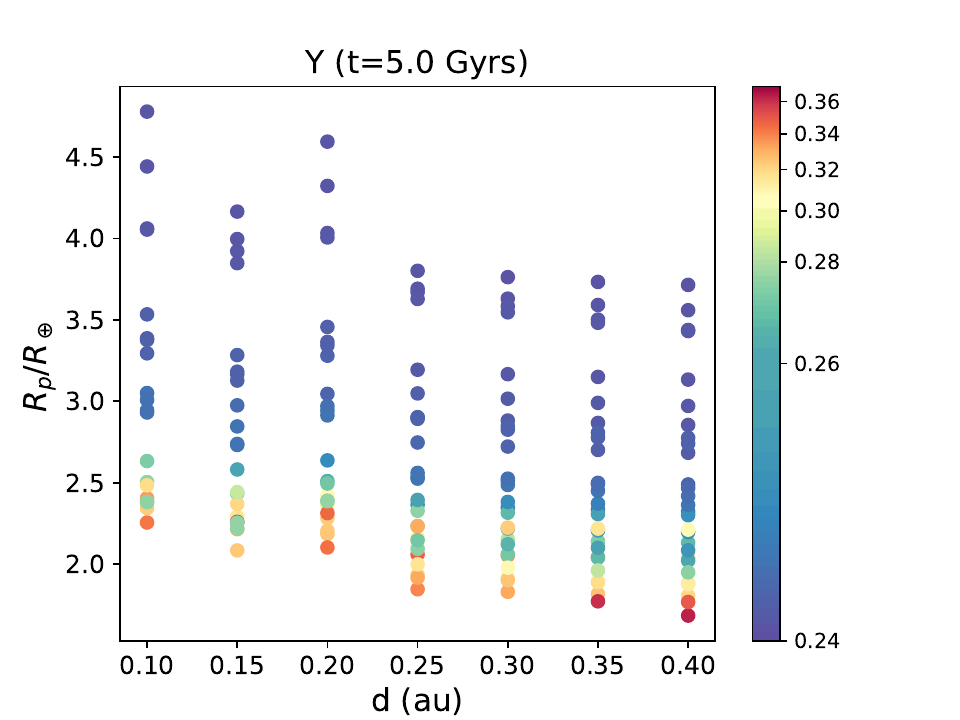}  &
\includegraphics[width=0.35\linewidth]{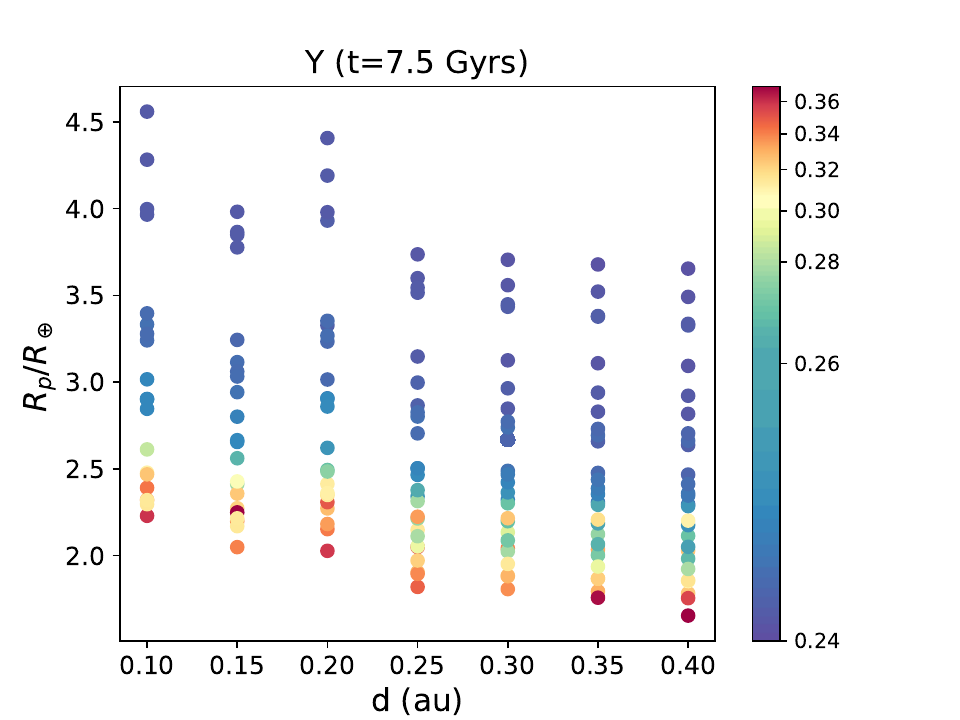} \\
\includegraphics[width=0.35\linewidth]{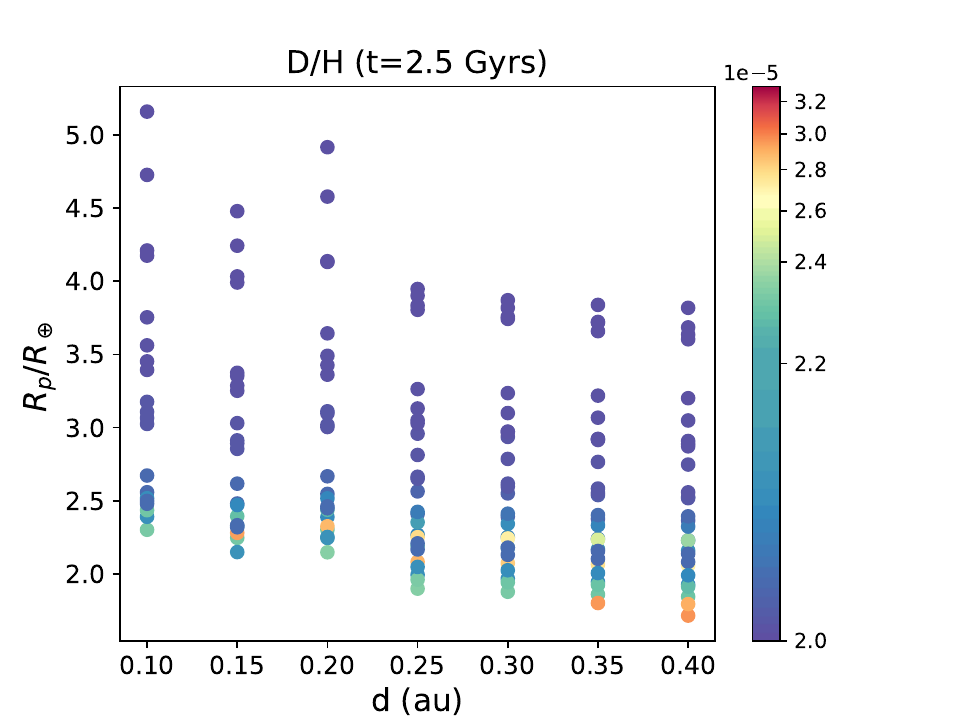} &
\includegraphics[width=0.35\linewidth]{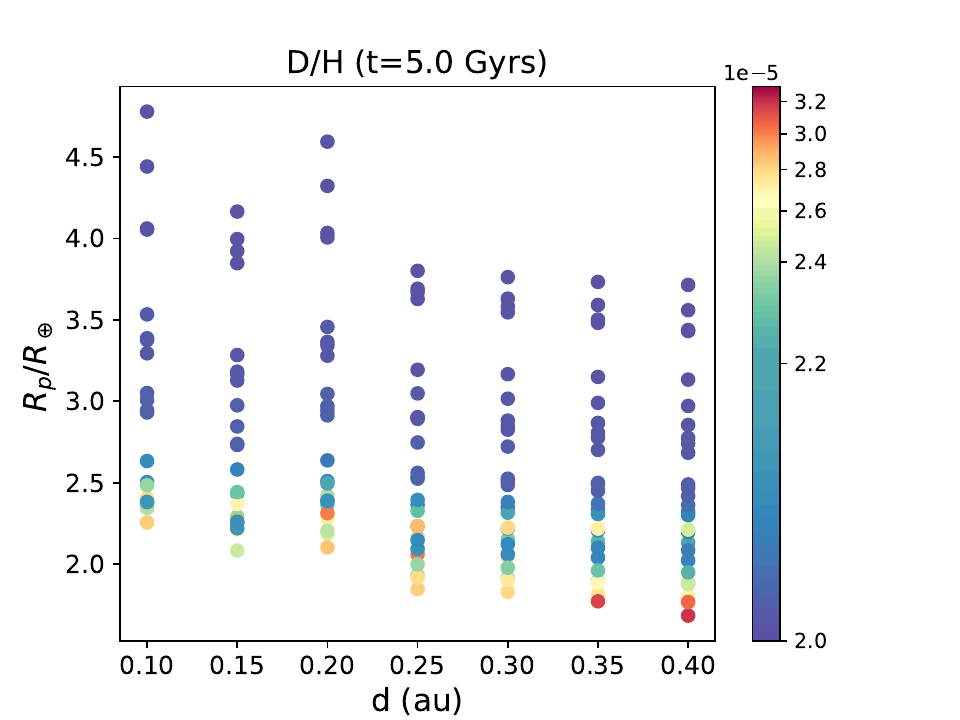}  &
\includegraphics[width=0.35\linewidth]{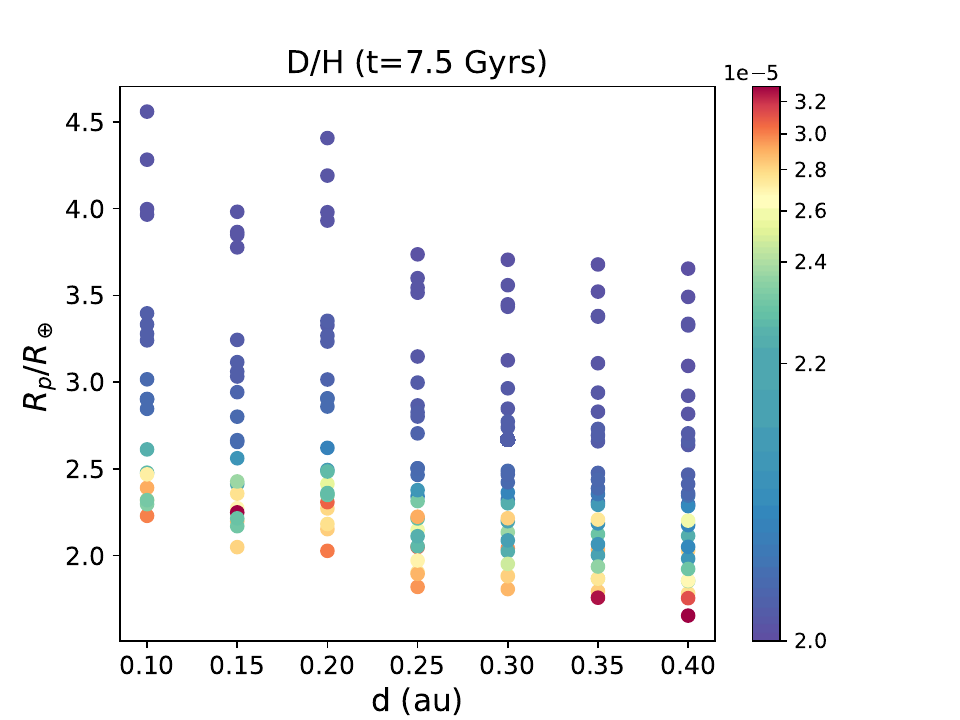}
\end{tabular}
\caption{Co-evolution of Y (top panels), D/H ratio (bottom panels), and $R_p$ (color coding) for a grid of planetary models: $M_p=$4, 5, 10, 15 $M_\oplus$ and $f_{env}=$0.001, 0.003, 0.005, 0.1, 0.03, 0.05, 0.1
at each orbital distance given by $d=$0.1, 0.15, 0.2, 0.25, 0.3, 0.35, and 0.4 au. Some data with anomalous evolution of planet radius are removed (see text). Similar to the evolution of helium-enhanced planets (top panels), the D/H ratios increase with time due to atmospheric escape and have larger values for the photoevaporating sub-Neptunes along the upper edge of the radius valley (i.e., $R_p \approx 1.8-2 R_\oplus$).}
\label{fig:grid}
\end{figure}

\subsection{A grid of simulations}
\label{sec:grid}

After understanding the basic outcomes of the H-D-He fractionation driven by EUV photoevaporation, we perform a grid of  simulations in a parameter space covering the planet's mass $=$(4, 5, 10, 15) $M_\oplus$, the orbital distance $d=$(0.1, 0.15, 0.2, 0.25, 0.3, 0.35, 0.4) au, and $f_{env}=$(0.001, 0.003, 0.005, 0.1, 0.03, 0.05, 0.1). 
We find that in the presence of fractionation, either a
large escape rate (i.e., small $d$) or a thin envelope (i.e., small $f_{env}$) of a low-mass planet somehow causes the atmosphere to contract abruptly and thus unphysically due to the evolution to a wrong equation of state in the MESA code. The data with the anomalous evolution of planet radius are removed. Consequently, there are no planets with bare rocky cores in our results. Nevertheless, the fate of these planets with anomalous size evolution could be either the planets with a very thin H-He atmosphere or bare cores. The parameter space and time frame for which the bare cores could be produced in the simulations will be discussed in Section~\ref{sec:dis1}.

We find that planetary H/He envelopes are more helium-abundant toward $R_p \approx 1.8$-$2R_\oplus$, particularly at later times.
This can be seen in Figure~\ref{fig:grid}, in which we show the results of the helium mass fraction Y and D/H ratio in the gas envelope of the model planets, along with the planet radius $R_p$, at $t=2.5$, 5, and 7.5 Gyr. This figure resembles Figure~5 of \citet{MR20} and Figure~1 of \citet{MR23} for the results of the Y evolution from their grids of planetary models. However, because we are not able to simulate a few cases with small $f_{evn}$, the Y values larger than 0.4 in the photoevaporating envelope,
as presented in \citet{MR23}, are not produced during the evolution from our grid of models. Nevertheless, Figure~\ref{fig:grid} shows that the
photoevaporating planetary envelopes are more helium-abundant toward $R_p \approx 1.8$-$2R_\oplus$ at later times, consistent with the result of \citet{MR23} that helium-enhanced planets lie along the upper edge of the radius valley where $R_p \approx 1.8$-$2 R_\oplus$.

Our calculated D/H ratios evolve in a similar manner to Y. This can be seen by comparing the top and bottom panels of Figures~\ref{fig:grid}, in which the gas envelopes of the planets with larger Y generally exhibit higher D/H ratios toward the upper edge of the radius valley at later times. It is also expected from the typical evolutions shown in Figure~\ref{fig:evol_0.2}. Specifically, the D/H ratio of the planets evolving to $R_p \approx 1.8$-$2R_\oplus$ can rise to $\approx 2.5$, 3.0, and 3.3 $\times 10^{-5}$, or increase by a factor of 1.25,  1.5, and 1.65, at the ages of 2.5, 5, and 7.5 Gyrs, respectively. As explained in the preceding paragraph, our grid of models limits the Y values because some cases with small values of $f_{env}$ are removed. Hence, the D/H ratio of the planets with a thinner atmosphere along the upper edge of the radius valley is expected higher than those shown in the bottom panels of Figure~\ref{fig:grid}, as long as these planets do not evolve to bare rocky cores.

\begin{figure}
\centering
\setlength{\tabcolsep}{0pt}
\begin{tabular}{ccc}
\includegraphics[width=0.35\linewidth]{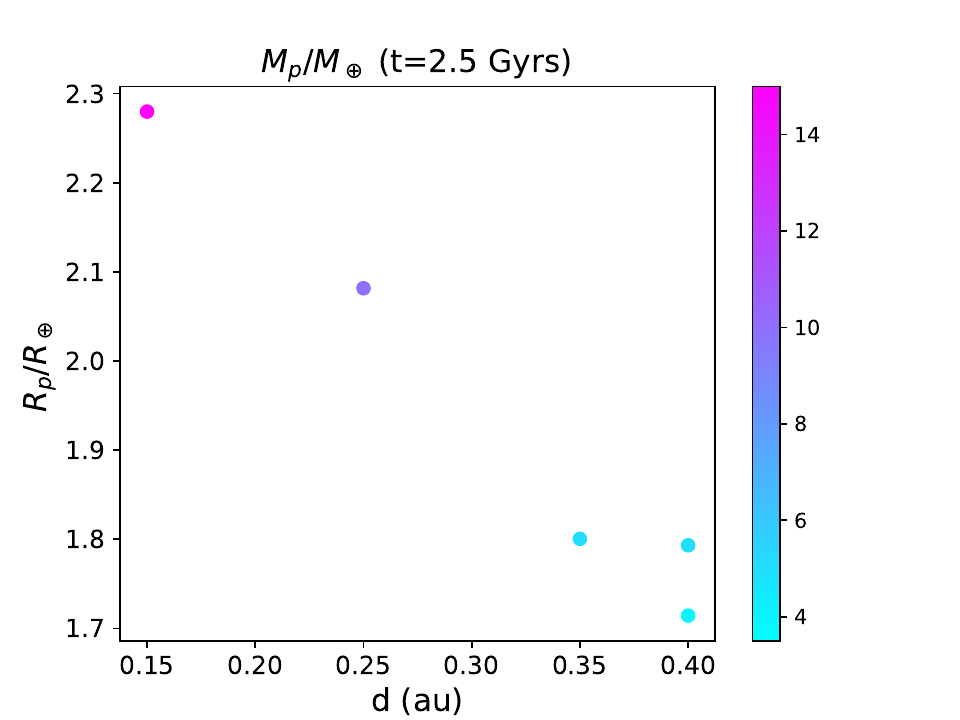} &
\includegraphics[width=0.35\linewidth]{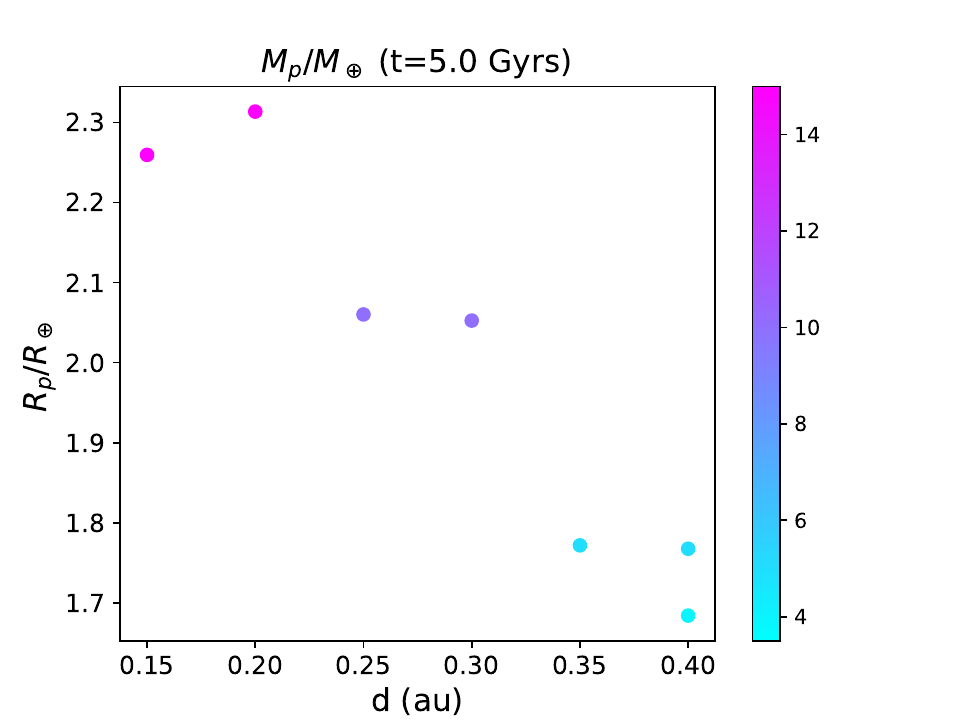}  &
\includegraphics[width=0.35\linewidth]{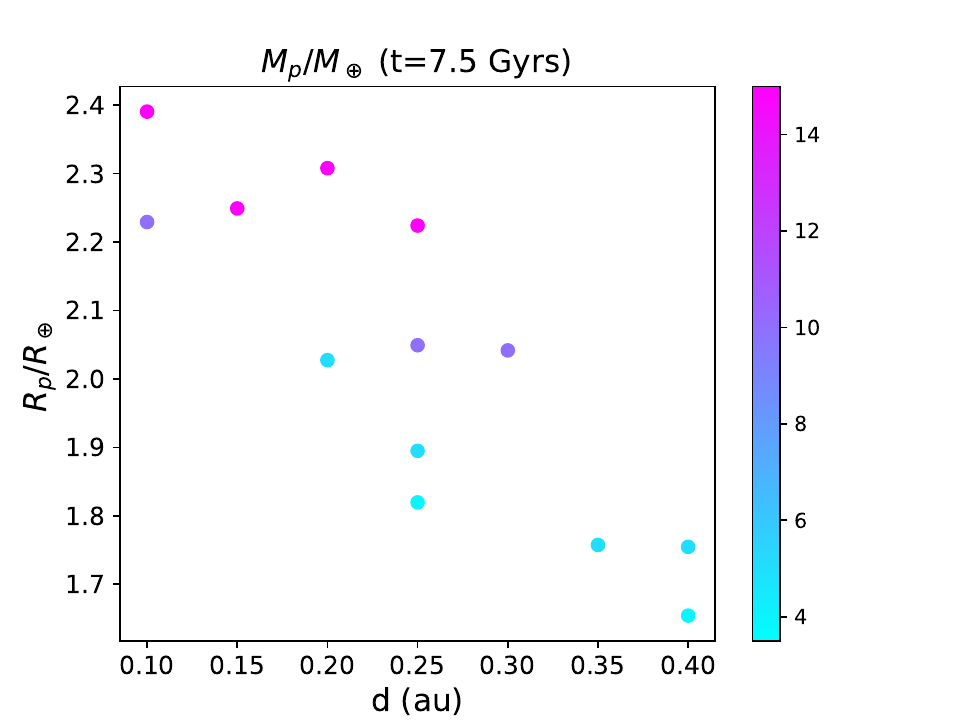} \\
\includegraphics[width=0.35\linewidth]{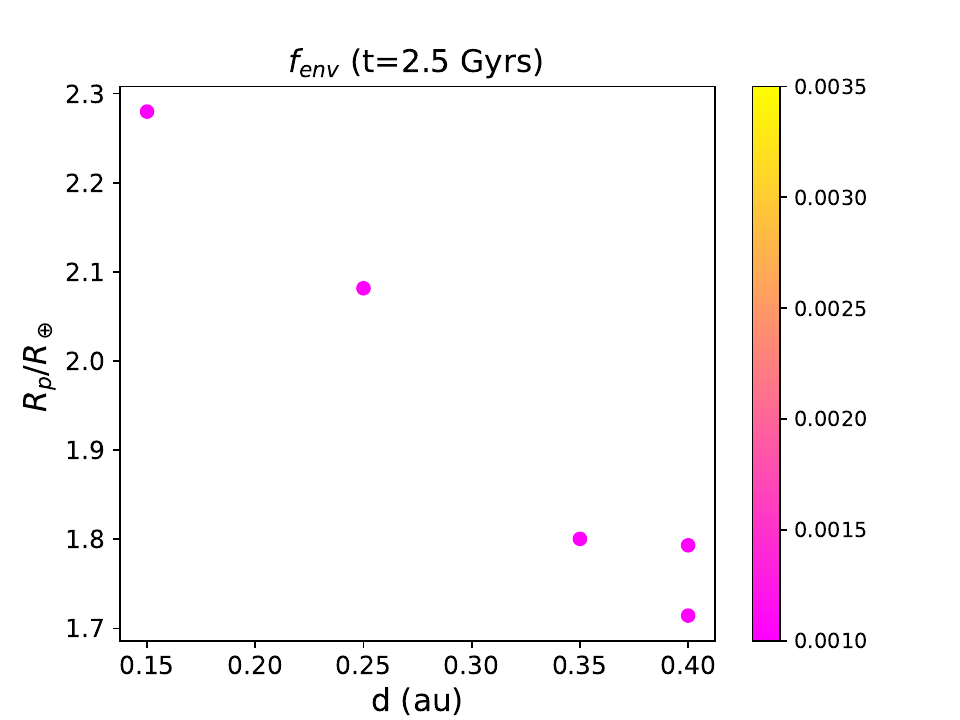} &
\includegraphics[width=0.35\linewidth]{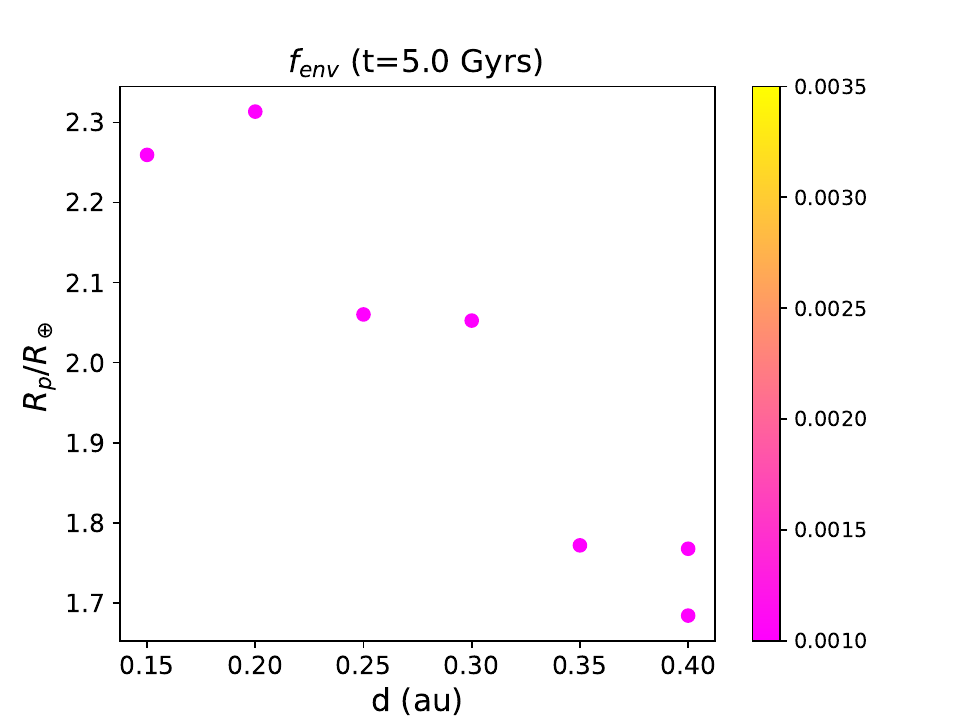}  &
\includegraphics[width=0.35\linewidth]{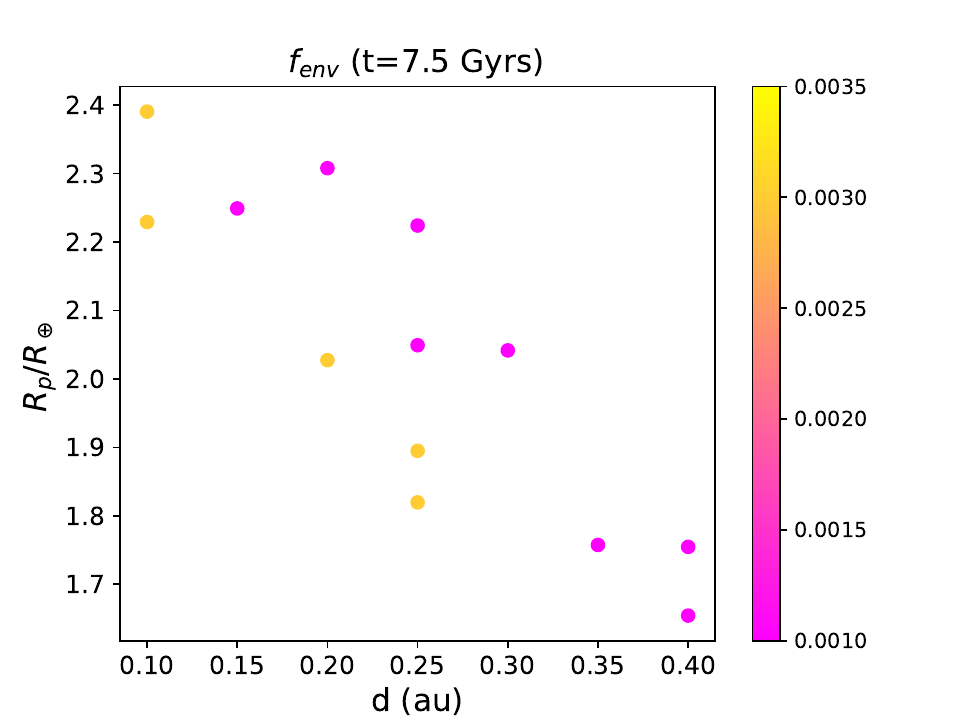}
\end{tabular}
\caption{The mass-radius-distance (top panels) and the envelope-radius-distance (bottom panels) parameter spaces for the deuterium-enhanced atmosphere in our grid of simulations shown in Figure~\ref{fig:grid}.
In the parameter space from $(R_p,\ d) \approx (2.4R_\oplus,\ 0.1$ au) to (1.7$R_\oplus$, 0.4 au), the data points
increase with time, meaning more deuterium-enhanced sub-Neptunes along the upper edge of the radius valley during the photoevaporation evolution. As illustrated in the bottom panels, the planets with initial low envelope fractions, i.e., $f_{env}=0.001$ and 0.003, can become deuterium-enhanced, which is expected from Figure~\ref{fig:evol_0.2}. The top panels show that more massive planets, i.e., $M_p=10$ and 15$M_\oplus$, with low $f_{env}$, can become deuterium-enhanced as the orbital distance decreases.}
\label{fig:grid_paramS}
\end{figure}

Our simulations result in more deuterium-enriched planets along the upper boundary of the radius valley, particularly those with lower $R_p$. 
These results are shown in Figure~\ref{fig:grid}. By including $M_p$ and $f_{env}$, we identify the parameter space for which greater
deuterium enhancement occurs in the planetary envelopes. 
Without the loss of generality, we define the D/H values $\geq 2.4\times 10^{-5}$ to be deuterium-enriched, which corresponds to an increase of 20\% and more in the ratio. The results are shown in Figure~\ref{fig:grid_paramS} for the $R_p$-$d$ parameter plane with color-coded values of $M_p$ (top panels) and $f_{env}$ (bottom panels). As expected, the data points for deuterium-enhanced planets lie along the lower boundary of the simulated results presented in Figure~\ref{fig:grid}. Furthermore, the bottom panels of Figure~\ref{fig:grid_paramS} show that these planets initially possess thin atmospheres, i.e., $f_{env}=0.001$ and 0.003, as expected from the typical evolution illustrated in Figure~\ref{fig:evol_0.2}. In other words, the sub-Neptunes with a higher $f_{env}$ become less deuterium-enhanced, which populate 
the upper part
of the data points in Figure~\ref{fig:grid} (i.e., larger $R_p$).
It is evident from Figure~\ref{fig:grid_paramS} that the deuterium-enhanced planets populate more widely in the parameter space of $f_{env}$-$R_p$-$d$ from 2.5 to 7.5 Gyr. This trend is expected; namely, more deuterium-enriched sub-Neptunes appear along the upper boundary of the radius valley during the photoevaporation evolution. Moreover, the top panels of Figure~\ref{fig:grid_paramS} show that more massive sub-Neptunes with low $f_{env}$ can become deuterium-enhanced as the orbital distance decreases from 0.4 to 0.1 au. It arises because although the incident EUV flux is stronger at smaller $d$, the deeper gravitational potential of massive planets of low $f_{env}$ (and thus small $R_p$) can reduce the mass-loss rate and hence facilitate the deuterium fractionation in the photoevaporating atmosphere. 
In summary, in our grid of simulations with a low envelope mass fraction of less than 0.005, a low-mass sub-Neptune (4-$5M_\oplus$) at $\approx$0.25-0.4 au or a high-mass sub-Neptune (10-$15M_\oplus$) at $\approx$0.1-0.25 au can increase the D/H values by 20\% or more over 7.5 Gyr.
Note that the parameter spaces in each time frame shown in Figure~\ref{fig:grid_paramS} are expected to be broader because some sub-Neptunes with small $f_{env}$ at small $d$ are disregarded in the analysis due to an unphysical radius anomaly in the simulations.


\section{Discussion}
\subsection{ Emergence of bare rocky cores and surviving sub-Neptunes}
\label{sec:dis1}
In our grid of simulations, the planets with rapid mass loss (i.e., lower $d$) and thin initial atmospheres (i.e., lower $f_{env}$) can go through a rapid contraction due to the incorrect equation of state in their evolution. 
As described in Section \ref{sec:grid},
we expect that the planets in some of our false simulations would lead to higher D/H ratios if they do not evolve to bare cores. It is difficult to further identify which planets would become bare in the simulations with incorrect evolution tracks. Nevertheless, we make an attempt to explore the possible parameter space and time frame for the emergence of bare cores in this subsection.
We note that a subset of these planets with anomalous radius evolution loses almost all the envelope and stops evolving, while the rest of them can continue to evolve up to $t=7.5$ Gyr at the end of the simulations.

Despite the numerical limitation for the simulations with fractionations,  
no issues associated with radius anomalies are found in the same grid of MESA simulations  in the absence of H-D-He fractionation. We find that the fates of the planets, bare rocky cores or surviving sub-Neptunes, are almost identical between the cases with and without fractionations, even though the simulations involving fractionations with radius anomaly are incorrect. Given the great similarities, we may use the results without fractionations to provide clues about the parameter space and time frame for the appearance of bare cores in the grid of simulations with fractionations.

Figure~\ref{fig:core} presents the three-dimensional parameter spaces for simulated results starting from the same initial conditions as those shown in Figure~\ref{fig:grid} but with evolution in the absence of fractionations. The sub-Neptunes shown by gray dots are almost identical to those presented in Figure~\ref{fig:grid} in terms of $R_p$ and $d$.
We only show the results with $R_p \leq 2.5 R_\oplus$ for clear visualization of the parameter spaces and time frames for the emergence of bare cores, which are illustrated by color-coded crosses.  Because no sub-Neptunes evolve to bare cores until $t\approx 1.8$ Gyr in the simulations, Figure~\ref{fig:core} presents the results in the time frames of 2.5, 5, and 7.5 Gyr, as shown in Figure~\ref{fig:grid}. 
It is clear that a radius valley appears between most bare cores and sub-Neptunes (i.e., $ 1.6 R_\oplus \lesssim R_p \lesssim 2R_\oplus$) and that the radius of the valley decreases with the orbital period ($d=0.1$-0.4 au corresponds to an orbital period of about 10-90 days). This location of the radius valley is generally consistent with 
the valley in the photoevaporation model by \citet[][see their Figure 5]{OW17}. A more detailed comparison is outside the scope of this study, as the radius valley also depends on photoevaporation models and the core mass distribution, which are not considered here (also see the next subsection for more discussion).\footnote{For instance, \citet{OW17} considered a Rayleigh distribution for the core mass, which peaks around 3 $M_\oplus$ with a standard deviation of 3 $M_\oplus$.}

From Figure~\ref{fig:core}, it is clear that more sub-Neptunes evolve to bare cores over gigayear timescales. Specifically,
the $M_p$-$R_p$-$d$ and $f_{env}$-$R_p$-$d$ parameter spaces of bare cores expand to larger $M_p$ (4-$10M_\oplus$), $R_p$ (1.5-$1.8R_\oplus$), $f_{env}$ (0.001-0.003), and $d$ (0.1-0.25 au) from 2.5 to 7.5 Gyr.  The bare cores emerge a bit late in comparison with the timescale for the high EUV flux of young Sun-like stars, which is about $10^8$ years (see Appendix \ref{sec:app2}). It arises due to the additional saturation factor $f_r$ in our photoevaporation model compared to the energy-limited models. 

On the other hand, the sub-Neptunes shown by green dots almost correspond to those removed from Figure~\ref{fig:grid} due to radius anomaly but with complete evolution over 7.5 Gyr. These planets populate the parameter spaces between those for bare cores (crosses) and those for sub-Neptunes that correspond to the planets shown in Figure~\ref{fig:grid} (gray dots). We suggest that the atmospheres of these surviving sub-Neptunes would potentially yield higher D/H values than those shown in the bottom panels of Figure~\ref{fig:grid}.

\begin{figure}
\centering
\setlength{\tabcolsep}{0pt}
\begin{tabular}{ccc}
\includegraphics[width=0.35\linewidth]{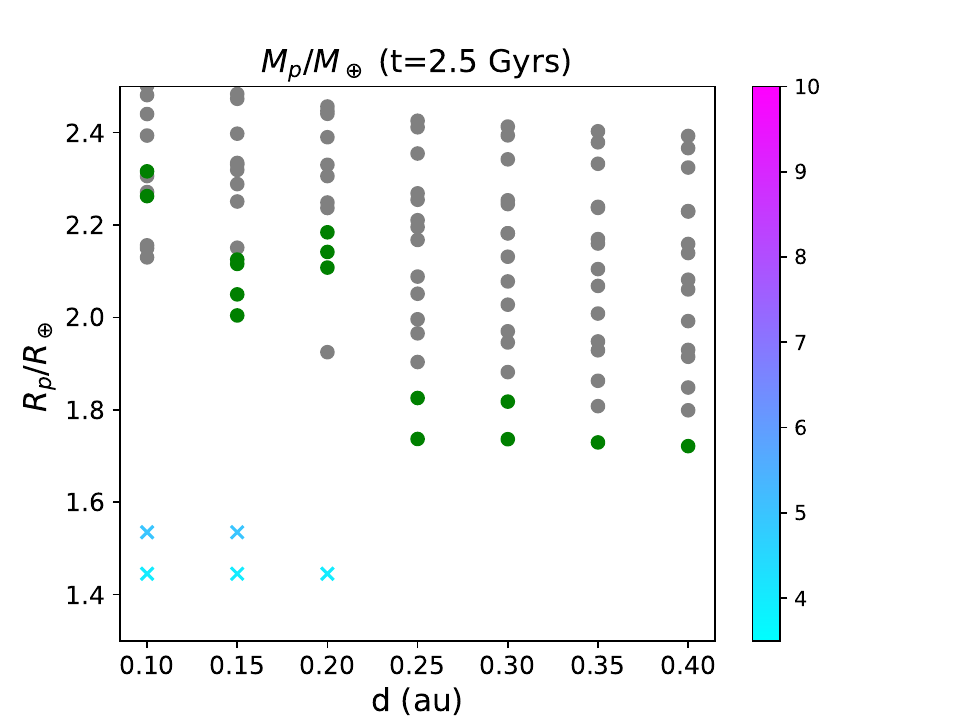} &
\includegraphics[width=0.35\linewidth]{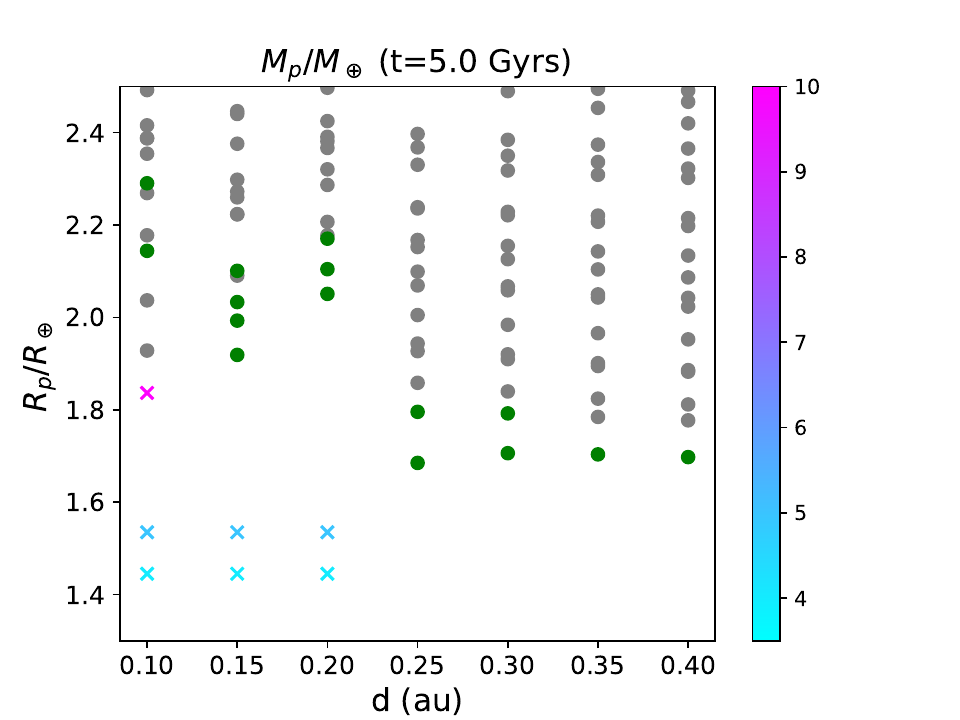}  &
\includegraphics[width=0.35\linewidth]{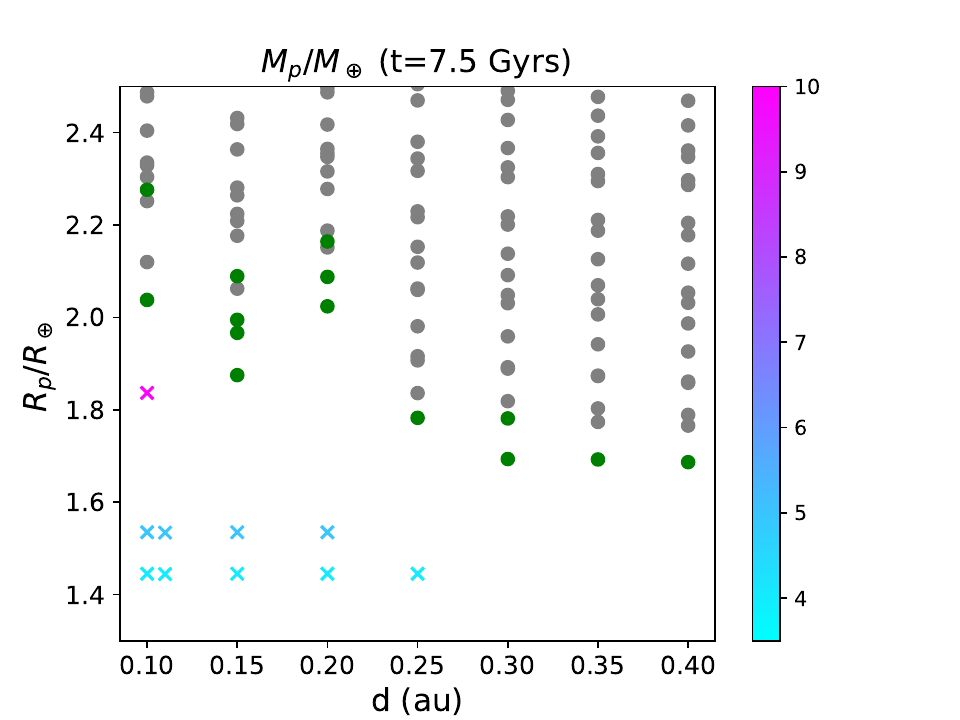} \\
\includegraphics[width=0.35\linewidth]{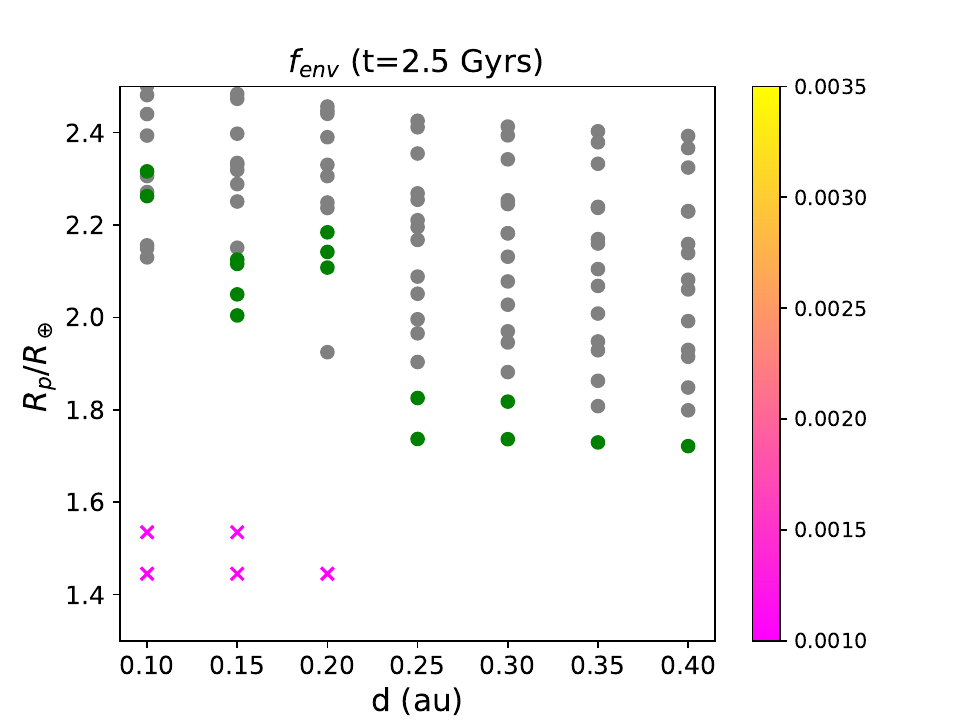} &
\includegraphics[width=0.35\linewidth]{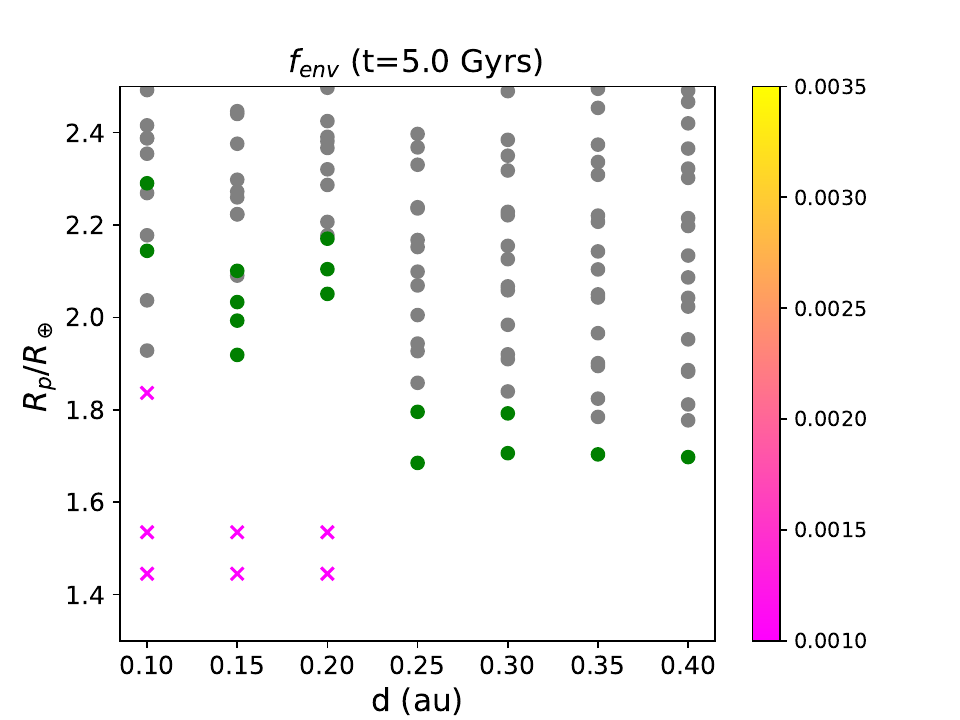} &
\includegraphics[width=0.35\linewidth]{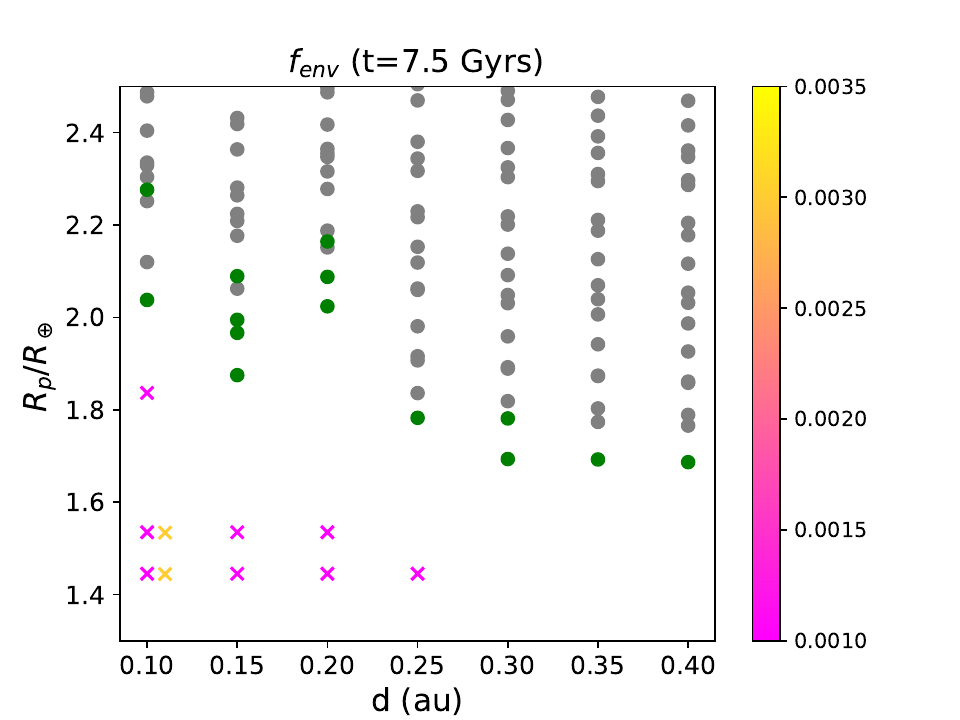} 
\end{tabular}
\caption{Mass-radius-distance (top panels) and envelope-radius-distance (bottom panels) parameter spaces and time frames for the appearance of bare rocky cores (denoted by color-coded crosses) in our grid of simulations without the H-D-He fractionation. The bare cores of the same $d$ and $R_p$ are plotted side by side to distinguish them. Despite no fractionations, the sub-Neptunes shown by gray dots are identical to those in Figure~\ref{fig:grid}. The sub-Neptunes in the $d$-$R_d$ parameter space shown by green dots survive up to $t=7.5$ Gyr in the grid of simulations without fractionations, whereas they are disregarded from Figure~\ref{fig:grid} due to radius anomaly in the grid of simulations with fractionations. The parameter spaces of bare cores become wider with time, meaning more sub-Neptunes evolving to bare cores over time.}
\label{fig:core}
\end{figure}

\subsection{Photoevaporation models}
In this study,  we include deuterium in the coupled thermal/mass-loss/compositional evolution of sub-Neptunes. However, our approach is similar to \citet{MR20} and \citet{MR23} in that 
an energy-limited approach for the photoevaporation is adopted, with the 
mass-loss rate suppressed by a direct simulation Monte Carlo (DSMC) factor of $f_r$ \citep[][also see Appendix \ref{sec:app2}]{Johnson,Hu15}.  
The photoevaporation efficiency $\eta$ in Equation(\ref{eq:Phi_EUV}) depends on the radiative cooling of the winds and is often taken as a constant of $\sim$1.5-4 \citep{Johnson}. In the energy-limited approach ignoring kinetic effects, $\eta$ can be modeled as a power law of the escape velocity at the planet photosphere \citep[e.g.,][]{OW17,Rogers21}. When the EUV flux from young host stars is high enough to allow for radiation recombination equilibrium in planetary winds, the escaping flow is limited by radiative cooling via Lyman-alpha lines \citep[e.g.,][]{Murray,CR16}. Recent hydrodynamical simulations suggested that $\eta$ can be expressed by an analytical function of gravitational potential and/or incident EUV flux to cover both energy- and radiation recombination-limited regimes of mass loss \citep{Salz,Caldiroli}. In Appendix \ref{sec:app3}, we show that our model with $\eta=0.08$-0.75 does not significantly change the results because the escape rate is limited by $f_r$ for the transonic flow driven by high EUV flux \citep{Hu15}. Therefore, we expect that applying the radiation recombination-limited model to our grid of simulations would not dramatically change the results of the H-D-He fraction. Future comparison studies based on the aforementioned photoevaporation models are worthwhile to test our expectation.

\citet{Kuby18a,Kuby18b} revised the photoevaporation rate based on a grid of hydrodynamic
models for hydrogen-only atmospheres. 
The authors demonstrated that the mass-loss rate based on the energy-limit model
is underestimated by several orders of magnitude in the regime of the low Jeans escape
parameter  $\Lambda$, which is a ratio of the gravitation energy to the intrinsic thermal energy for a hydrogen atom in an atmosphere.
In this regime, the escape is primarily caused by low gravity and high equilibrium temperature of planets and depends weakly on stellar EUV flux.
Hence, under the assumption that the revised model works properly for standard atmospheres containing both hydrogen and helium, the 
mass-loss rate of less massive planets should be drastically enhanced, resulting in different H-D-He fractionations in the atmospheres. Nevertheless, the observed bimodal size distribution and the radius valley for {\it Kepler} small planets can be reproduced with the revised photoevaporation model as long as the planet population shifts to a high-mass distribution \citep{MLM21,KP23}.  We then expect that the evolved D/H ratio would be similar to what we derive and still be highest along the upper edge of the radius valley if high-mass planets are considered. A study will be conducted to examine the speculation using the latest version of MESA to possibly cope with the issue of the equation of state in this fast-wind model.

Additionally, while our study focuses on solar-type stars with a fiducial surface temperature of 6000 K and adopts the EUV flux described in Appendix \ref{sec:app2}, it is noteworthy that the time-integrated X-ray exposure of a planet at a fixed incident bolometric flux decreases with stellar mass \citep{McDonald19}. To better understand the nature of the radius valley, this effect depending on the stellar mass was considered to distinguish the photoevaporation from the core-powered mass-loss model \citep{Rogers21,Berger23}. The stronger mass loss driven by more intense X-rays may be hostile to deuterium fractionation. Investigating the D/H ratio of planets around various stellar types in relation to the radius valley in multiple-parameter space is outside the scope of this work and will be studied in future work.

\subsection{Core-powered mass-loss model}
In our grid of planet models with the same initial entropy specified by \citet{MR20}, the initial temperature of the envelope base is a few thousand kelvins. If we assume that the entire core temperature is the same as that at the envelope base, our planet model would be consistent with the molten rocky core scenario to drive a significant atmospheric escape, i.e., core-powered mass loss  \citep{Ginzburg16, Ginzburg18}. As described in the Introduction, this alternative model also successfully produces the observed radius valley \citep{GS19,Berger23}.
When a low-mass planet is young and thus large such that $\Lambda$ is small, the core-powered mass loss of gas envelope can be much more substantial than the classic energy-limited photoevaporation \citep{Kuby2021}. The core-powered mass loss arises from the slow release of the core internal energy through the optically thick envelope on $\sim$ Gyr timescales \citep[e.g.,][]{GS20,Rogers21}. Although heating due to radioactive decay in the core has been implemented in our MESA module for photoevaporative escape \citep{CR16},
core-powered mass loss is not included in our current model. It remains to be examined whether the escape rate of the prolonged mass loss at later phases can decline close to the critical diffusive escape rate of deuterium $\Phi_{crit,D}$ to enable significant fractionation. In principle, both photoevaporation and core-powered mass-loss mechanisms are expected to occur during the lifetime of a sub-Neptune \citep{MK23}.

\subsection{Deuterium homopause}
In this work, $r_0$ in Equation(\ref{eq2_1:D}) is simply taken to be $R_h$, which is determined by the homopause of the hydrogen-helium mixture, i.e., $K_{zz}=\mathcal{D}_{H,He}$. In principle, the deuterium homopause does not necessarily lie at the helium homopause in a hydrogen-rich atmosphere. Specifically,
since $X_D \ll X_H$ and $X_{He}$, we can consider D in the binary mixture of H and He and calculate the mixing diffusivity of deuterium in the mixture as follows \citep{Tang}:
\begin{equation}
{1\over \mathcal{D}_{D, H, He}} ={X_H \over  \mathcal{D}_{D,H}}+{X_{He} \over \mathcal{D}_{D,He}}.
\label{eq:D_3}
\end{equation}
Hence, the location of the deuterium homopause is determined by the relation $K_{zz}=\mathcal{D}_{D,H,He}$ rather than $K_{zz}=\mathcal{D}_{H,He}$.
With Equation(\ref{eq:D_3}), we can estimate the difference between $\mathcal{D}_{D,H,He}$ and $\mathcal{D}_{H,He}$ at a given altitude (i.e. given temperature and pressure) to assess our simplified approach.  To estimate $ \mathcal{D}_{D,H}$ and  $ \mathcal{D}_{D,He}$ in  Equation(\ref{eq:D_3}), the atomic diffusion volume of deuterium $V_D$ is needed (see Equation(\ref{eq:diffusivity})), which is unknown. We expect that the value of $V_D$ lies between those of $V_H$ and $V_{He}$.  It then follows that at a given altitude, $\mathcal{D}_{D,H,He}/\mathcal{D}_{H,He}$ is about 1.1 and 0.98 for $V_D=V_H$ and $V_{He}$, respectively; namely, the two binary diffusivities are almost identical. Furthermore, the H-He fractionation hardly changes when the eddy diffusivity $K_{zz}$ alters from $10^7$-$10^{11}$ cm$^2$/s \citep{MR20}, and the same consequence applies to the H-D fractionation in our model (see Appendix \ref{sec:app3}). 
All of the above reasons suggest that considering $R_h$ as the homopause for deuterium is reasonable in our problem. 

\subsection{Observational prospects}

Here we describe some potential observational avenues to test the suggested D/H signatures.
Assuming that the initial sub-Neptunes in the solar neighborhood are close to the protosolar abundance, i.e., D/H$\approx 2\times10^{-5}$ and CH$_3$D/CH$_4\approx 8\times 10^{-5}$, it would be possible for future ground-based telescopes, such as the Extremely Large Telescope, using the cross-correlation technique \citep{MS19} to probe vibrational emissions from deuterated methane CH$_3$D at $\sim 4.7 \mu$m toward these planets with equilibrium temperatures below 600 K (i.e. roughly corresponding to $d>0.2$ au in this study).

Previous work sampling a variety of Earth-like abundance isotopologues on Venus-like atmospheres has shown that fractionation signals from species such as HDO can be accomplished with {\it JWST}'s NIRSpec in as few as 10 transits \citep{Lincowski19}. Similarly, infrared spectroscopy on the atmospheres of sub-Neptune-sized planets should aim to characterize HDO features at 3.7, 2.4, and 1.5 $\mu$m. Despite the lower degrees of fractionation in our simulations (i.e., for a Venus-like atmosphere, D/H in water vapor is assumed to be $\sim 100 \times$ (D/H)$_{\rm ocean}$) compared to our case of $\gtrsim 2 \times 1.65 \times$ (D/H)$_{\rm protosolar}$), the much larger scale heights of extended sub-Neptune planet envelopes might aid the detection of deuterium and other species associated with photoevaporation-induced fractionation, particularly for the strongly irradiated cases (e.g., $d = 0.1$ au, Figure~\ref{fig:grid}).
An investigation into the detectability of deuterated water through the transmission spectra of nearby sub-Neptunes would be encouraged to constrain our model.
While significant deuterium fractionation could indicate a more evolved atmosphere, even a non-detection of deuterium enhancement may be attributed to planet age, other unknown assumptions (such as mass-loss efficiency, initial disk D/H, and metallicity), or additional considerations not included in this work (such as nonthermal escape, inclusion of interactive chemistry, and the possibility of deep H$_2$O reservoirs).


\section{Summary}
We further extend the MESA module based on the EUV photoevaporation model adopted by \citet{Hu15} and \citet{MR20}.  For the first time, we conduct compositional simulations coupled with the thermal interior structure to study the evolution of D/H of photoevaporating sub-Neptunes. The key results are summarized as follows.
\begin{itemize}
 \item We derive the critical diffusive escape rate for deuterium $\Phi_{crit,D}$ in a hydrogen- and helium-rich atmosphere. The deuterium ceases to escape when the atmospheric escape rate becomes smaller than $\Phi_{crit,D}$.
 \item  The planets with smaller $f_{env}$ and thus smaller $R_p$ can lead to smaller $\dot M_p$, which facilitates fractionation. 
 Specifically, in our grid of simulations with low $f_{env}<0.005$, a low-mass sub-Neptune (4-$5M_\oplus$) at $d\approx$ 0.25-0.4 au or a high-mass sub-Neptune (10-$15M_\oplus$) at $d\approx$ 0.1-0.25 au can increase the D/H values by 20\% or more over 7.5 Gyr.
 \item Analogous to the helium-enhanced planets \citep{MR23}, the planets along the upper boundary of the radius valley are the best targets to detect high D/H ratios in their thin atmospheres.  The ratio can rise by a factor of $\lesssim$ 1.65 within 7.5 Gyr in our grid of evolutionary calculations, independent of the initial D/H value.
 \item A few cases of small $f_{env}$ are removed in our grid of simulations due to the numerical limitation. 
 Therefore, the D/H ratio is expected to increase more from the initial value. Assuming the initial D/H in the gas envelope of sub-Neptunes is protosolar, Figure~\ref{fig:DHcomp} illustrates the D/H value of the planets along the radius valley from our study in comparison with those in Earth's ocean, as well as in the gas and icy planets in the Solar System. 
\end{itemize}

\begin{figure}
\plotone{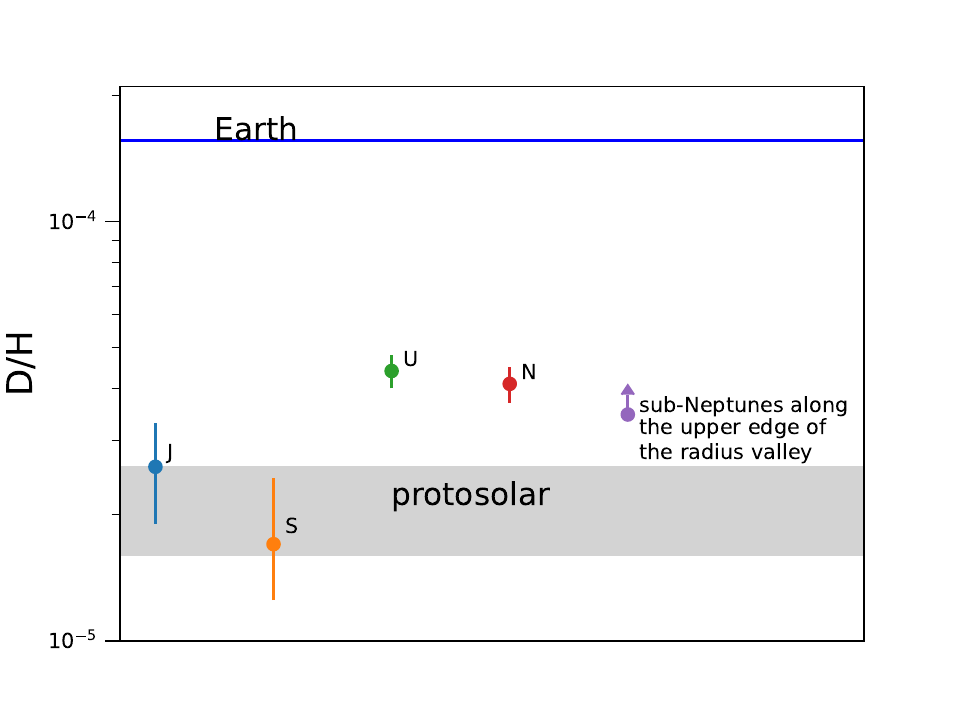}
\caption{The D/H ratio of giant and icy planets in the Solar System (J: Jupiter, S: Saturn, U: Uranus, and N: Neptune), along with the simulated ratio of the photoevaporating sub-Neptunes along the upper edge of the radius valley. 
The D/H values of Earth's ocean and the protosolar nebula are shown by blue and gray horizontal lines, respectively. Analogous to the helium-enhanced planets due to atmospheric escape \citep{MR23}, the planets along the upper boundary of the radius valley generally exhibit the largest D/H value in their thin atmospheres (see Figure~\ref{fig:grid}). The simulated D/H value along the upper radius valley only provides the lower limit.}
\label{fig:DHcomp}
\end{figure}




\acknowledgments
We thank Issac Malsky, Peter Bodenheimer, James Owen, and Teppei Okumura for the informative discussions. We also thank the anonymous referee for helpful comments that improved the quality of the manuscript. P.-G.G. acknowledges support from the National Science and Technology Council in Taiwan through grants NSTC 111-2112-M-001-037 and 112-2112-M-001-035.







\appendix
\section{Fractionation due to H-rich atmospheric escape}
\label{sec:app1}
We adopt the assumption that the lowermost subsonic regions of the wind control differential escape \citep{Zahnle86,Zahnle}. Therefore, in
a great depth of planetary winds where the flow is subsonic, its kinetic energy can be ignored for our interest in species fractionation. Under this circumstance,
the equation of motion for an isothermal wind for the species $j$ in a steady state, with H as the major constituent, may be approximated to 
\citep{Zahnle}
\begin{equation}
{1\over X_j}{d X_j \over dr}=-{GM_p (m_j- m_H) \over kTr^2} + \sum_i {r_0^2 \over r^2 b_{i,H}} \left( {X_i \over X_H}\phi_H
-\phi_i \right) + \sum_i {r_0^2 \over r^2 b{i,j}} \left( \phi_i - {X_i \over X_j}\phi_j \right),
\label{eq:orig}
\end{equation}
where $X$ is the mixing ratio,\footnote{Note that $X$ in this paper is defined as the mixing ratio to the entire gas mixture, while it is defined as the mixing ratio relative to the H abundance in \citet{Zahnle}.}  and $r$ is the radial coordinate. In the above equation, the species $i$ exerts a drag force on the flow of the species $j$, characterized by the binary diffusion coefficient $b$ defined by $kT/(\mu_{ij}k_{ij})$, where $\mu_{ij}$ is the reduced mass, and $k_{ij}$ is the collision rate.
For a non-negligible flow for the species $j$, its abundance would be much more uniform over one density scale height of H, i.e., $d \ln X_j/dr \ll d \ln n_H/dr$. It follows from the derivation of Equation(\ref{eq:orig}) that the term $d \ln X_j/dr$ in the above equation may be
ignored to allow for a simple analytical solution.

In the escaping flow consisting only of H and He from a photoevaporating planet envelope, the number fluxes of H and He are governed by mass and momentum conservation evaluated at the homopause $r_0=R_h$ \citep{Hu15,MR20,MR23}:
\begin{eqnarray}
\Phi &\approx &\Phi_H + \Phi_{He}=4 \pi R_h^2 (\phi_H m_H + \phi_{He} m_{He}), \label{eq:Phi}\\
{\phi_{He} \over X_{He}} &= & {\phi_H \over X_H}-\phi_{DL,He}, \label{eq:He}
\end{eqnarray}
Equation(\ref{eq:He}) is derived from Equation(\ref{eq:orig}) by neglecting $d \ln X_{He}/dr$, using $X_H+X_{He} \approx 1$, and employing the diffusion-limited escape flux for He given by
\begin{equation}
\phi_{DL,He}={GM_p(m_{He} -m_H) b'_{H,He} \over R_h^2 k T_h},
\label{eq:DL_He}
\end{equation}
where $T_h$ is the homopause temperature, $b'_{H,He}$ is the binary diffusion coefficient for H and He taking into account the H ionization.
Moreover, $R_h$ for H and He is determined by the location where the binary diffusivity $\mathcal{D}_{H,He}$ equals the eddy diffusivity $K_{zz}$, for which a nominal value $10^9$ cm$^2$/s adopted by \citet{MR20} is used.  $\mathcal{D}_{H,He}$  is determined by Fuller's method \citep{Fuller,Tang,MR20}:
\begin{equation}
\mathcal{D}_{H,He} ={10^{-3} T_h^{1.75} \left( {m_H m_{He} \over m_H + m_{He}} \right)^{-1/2} \over P_h \left( V_H^{1/3} +V_{He}^{1/3} \right)^2},
\label{eq:diffusivity}
\end{equation}
where $P_h$ is the gas pressure in units of atm at $R_h$, and $V_H (=1.98)$ and $V_{He} (=2.88)$ are the atomic diffusion volumes. 

Once $P_h$ is identified, the homopause radius $R_h$ is obtained in a hydrostatic atmosphere using the gas pressure, as well as constant values for gravitational acceleration $g$, molecular weight $\mu$, and the pressure scale height at MESA's outermost zone where the optical depth is $2/3$ (i.e., photosphere). Moreover, we use the transit radius, defined by a pressure level of 1 mbar, as the planetary radius $R_p$. Detailed descriptions of these radii and atmospheric boundary conditions for an irradiated sub-Neptune can be found in \citet{MR20} and \citet{MR23}.

\section{Model for EUV photoevaporation}
\label{sec:app2}
We adopt the mass-loss rate due to EUV photoevaporation given by \citet{Hu15}, \citet{MR20}, and \citet{MR23}
\begin{equation}
\Phi = f_r \Phi_{EL},
\label{eq:Phi_EUV}
\end{equation}
where 
$\Phi_{EL}=L_{EUV} \eta a^2 R_h^3 /(4K d^2 GM_p)$ is the energy-limited mass-loss rate for a planet at an orbital distance $d$. Here $L_{EUV}$ is the host-star EUV luminosity evolving with the stellar age $\tau$ described by the power law $\log_{10} (L_{EUV}/{\rm J\,s^{-1}})=22.12-1.24\log_{10} (\tau/{\rm Gyr})$ for GK dwarfs at an age $t\geq 10^8$ years \citep{Sanz11}. When $t< 10^8$ years, the stellar high-energy flux is almost saturated \citep[][references therein]{OW17}. Hence, $L_{EUV}$ for $t<10^8$ years is set to be the same as the power law of $L_{EUV}$ at a stellar age of $10^8$ years. Moreover in Equation(\ref{eq:Phi_EUV}), the heating efficiency $\eta=0.1$, the fraction of heating radius $a=1$,  and the Roche potential reduction factor $K$ are taken to be the same values and expression as those in \citet{Hu15} and \citet{MR20}. 
The $f_r$ is introduced in Equation(\ref{eq:Phi_EUV}) to recover the neglected terms associated with the kinetic energy and $d \ln X_j/dr$ (thermal energy) according to the DSMC, which corrects the overestimate of $\Phi$ due solely to $\Phi_{EL}$ \citep{Johnson,Hu15}.

\section{Sensitivity of D/H to free parameters}
\label{sec:app3}
In our fiducial model, the free parameters $\eta=0.1$, $K_{zz}=10^9$ cm$^2$/s, and $T_h=10000$ K are adopted to 
study the D/H evolution of close-in sub-Neptunes due to EUV photoevaporation. In this section, we investigate whether the D/H values are sensitive to these free parameters. Besides the fiducial values, we follow \citet{MR20} and \citet{MR23} and consider the eddy diffusion coefficient $K_{zz}=10^9$ and $10^{11}$ cm$^2$/s, and homopause temperature $T_h=3000$ K. We also consider $\eta=0.08$ and $0.75$ for the low and high photoevaporation efficiency, respectively.  The investigation is carried out such that the value of one parameter is changed at a time, while everything else remains the same as the fiducial values in each simulation. Figure~\ref{fig:freeparam} shows the resulting D/H evolution (dashed and dotted curves) compared with the typical evolution shown in Figure~\ref{fig:evol_0.2} (solid curves). 

Overall, the D/H values are insensitive to these free parameters, except in extreme cases for photoevaporation with $\eta=0.75$. For the influence of $K_{zz}$ and $T_h$, it has been shown that the mass-loss rate increases with $K_{zz}$ \citep{MR20} and the level of H-He fractionation in the escaping wind decreases with $T_h$ \citep{MR23}. These effects apply to H-D fractionation too, resulting in the outcome that the D/H values slightly increase with $K_{zz}$ and decrease with $T_h$, as shown in Figure~\ref{fig:freeparam}. However, the influences on the D/H values are insignificant.

On the other hand, the models with $\eta=0.75$ yield a much faster increase in D/H than those in the fiducial model in late evolution, as shown in the lower left panel of Figure~\ref{fig:freeparam} for the more massive sub-Neptune of 15$M_\oplus$ with smaller initial envelope mass fractions, e.g., $f_{env}=0.005$. What happens is that the 
mass-loss rates in most cases are approximately similar, even when the mass loss is significant due to the suppression factor $f_r$ \citep{Hu15}. For a massive sub-Neptune with a small $f_{env}$, the mass-loss rate starts to drop after $t\sim 1$ Gyr due to the weak EUV flux. Hence, $f_r$ becomes unity; thus, the magnitude of $\eta$  becomes influential during the late stage.  When $\eta=0.08$ and 0.1, deuterium can stop escaping during this late phase due to $\Phi < \Phi_{D,crit}$, hence increasing the D/H value (see the lower left panel of Figure~\ref{fig:evol_0.2}). However, this increase is still less than the rise of D/H ratio due to the faster diffusive escapes of both D and H when $\eta=0.75$ even though $\Phi$ has not reached $\Phi_{D,crit}$. This effect leads to the slightly large departure of the D/H values among different $\eta$ in the cases of $f_{env}=0.005$ and 0.1 at $t \gtrsim 2$-3 Gyr as shown in the lower left panel of Figure~\ref{fig:freeparam}. Despite this subtlety, we conduct the same grid of simulations and find that the largest D/H value in the models with $\eta=0.75$ and 0.08 is similar to that with $\eta=0.1$. This may infer that our results are insensitive to $\eta=0.08$(low)-0.75(high photoevaporation efficiency).  With that being said, we caution that the parameter space of the cases with the various values of $\eta$ is different due to numerical limitations. For example, as shown in the upper left panel of Figure~\ref{fig:freeparam}, the case for the set of free parameters given by $M_p=5M_\oplus$, $f_{env}=0.005$, and $\eta=0.75$ is disregarded due to the unphysical radius anomaly.

\begin{figure}
\centering
\setlength{\tabcolsep}{0pt}
\begin{tabular}{ccc}
\includegraphics[width=0.35\linewidth]{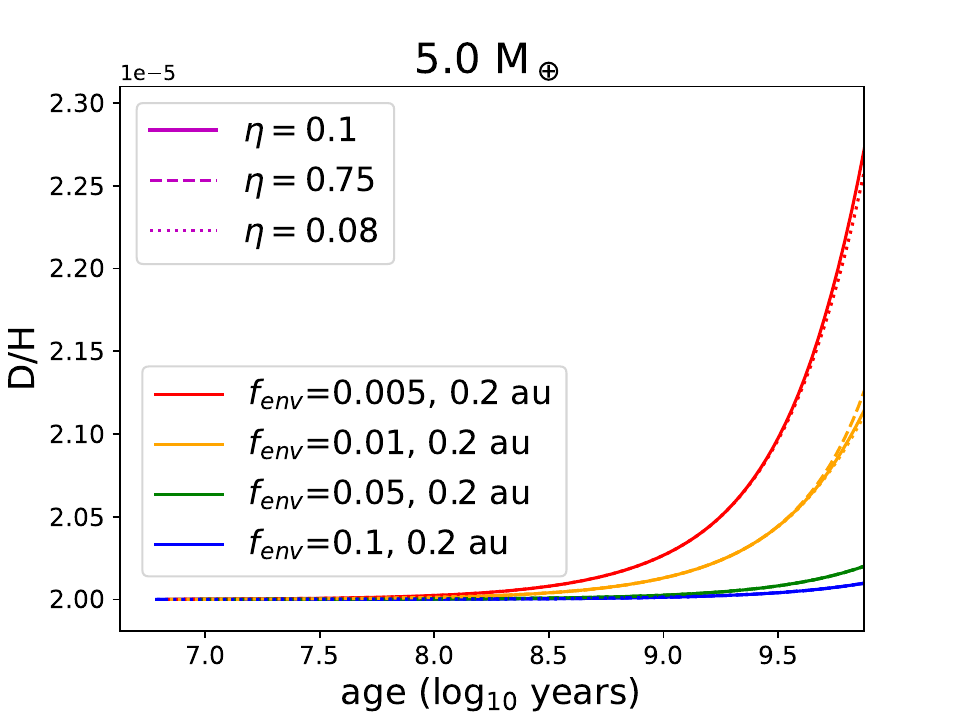} &
\includegraphics[width=0.35\linewidth]{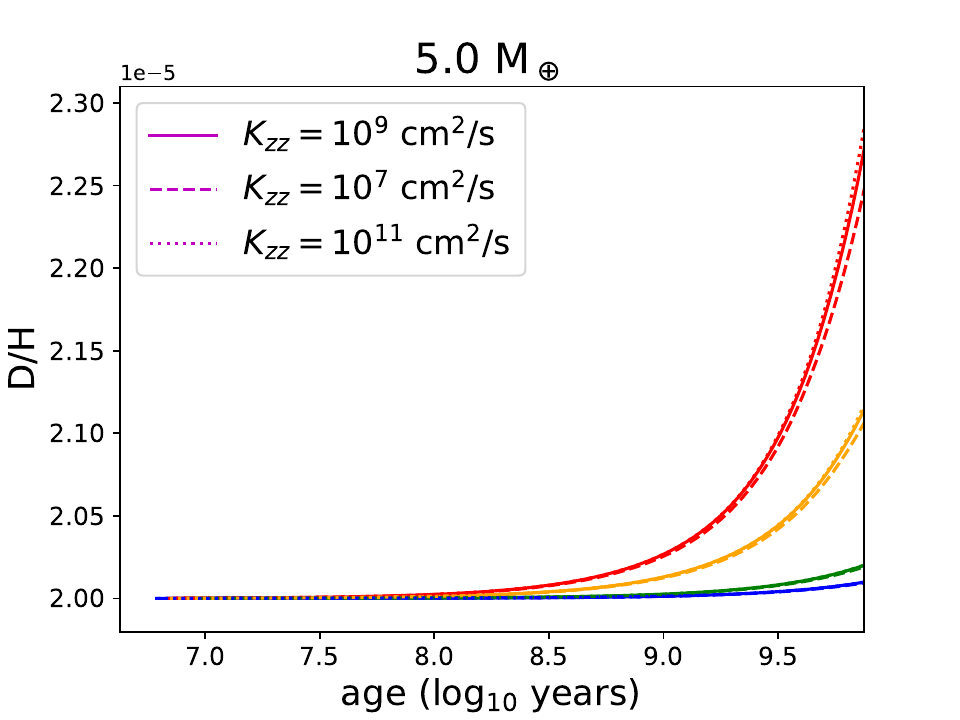}  &
\includegraphics[width=0.35\linewidth]{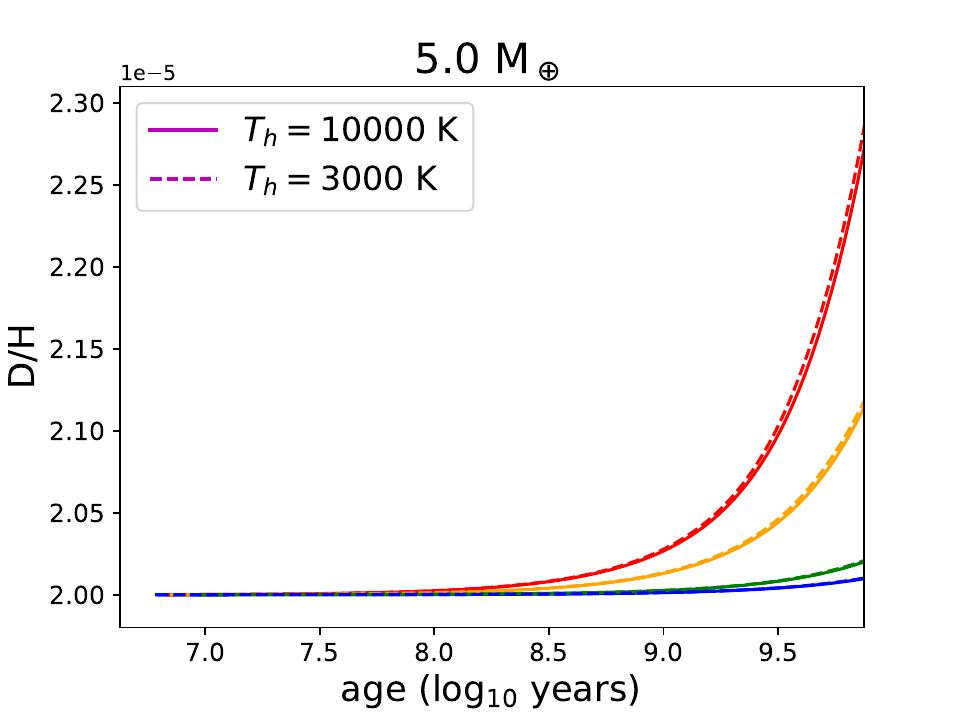} \\
\includegraphics[width=0.35\linewidth]{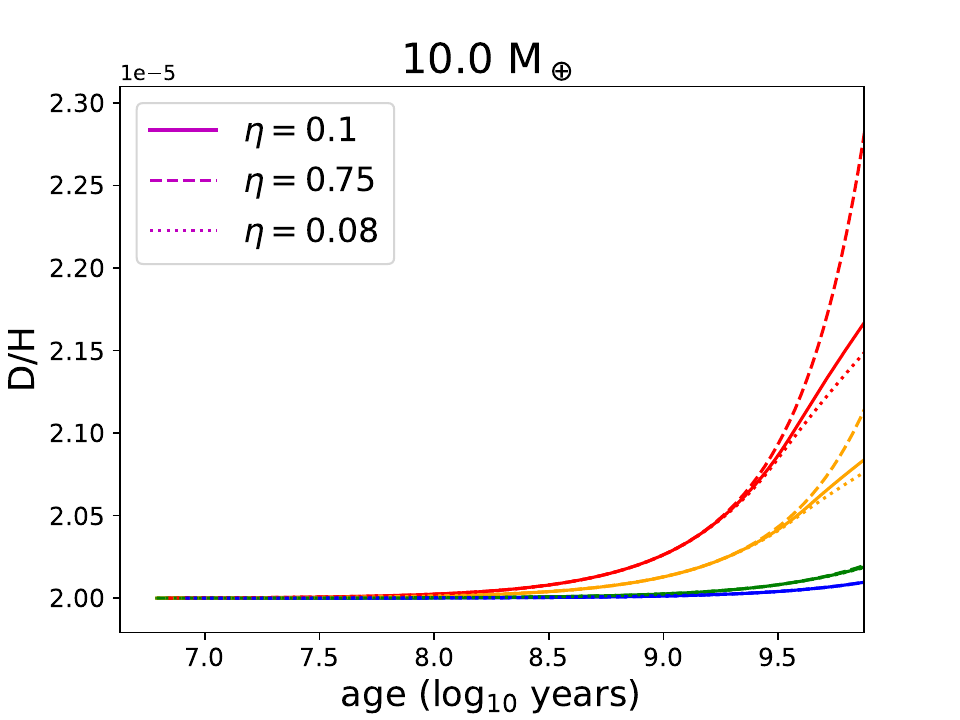} &
\includegraphics[width=0.35\linewidth]{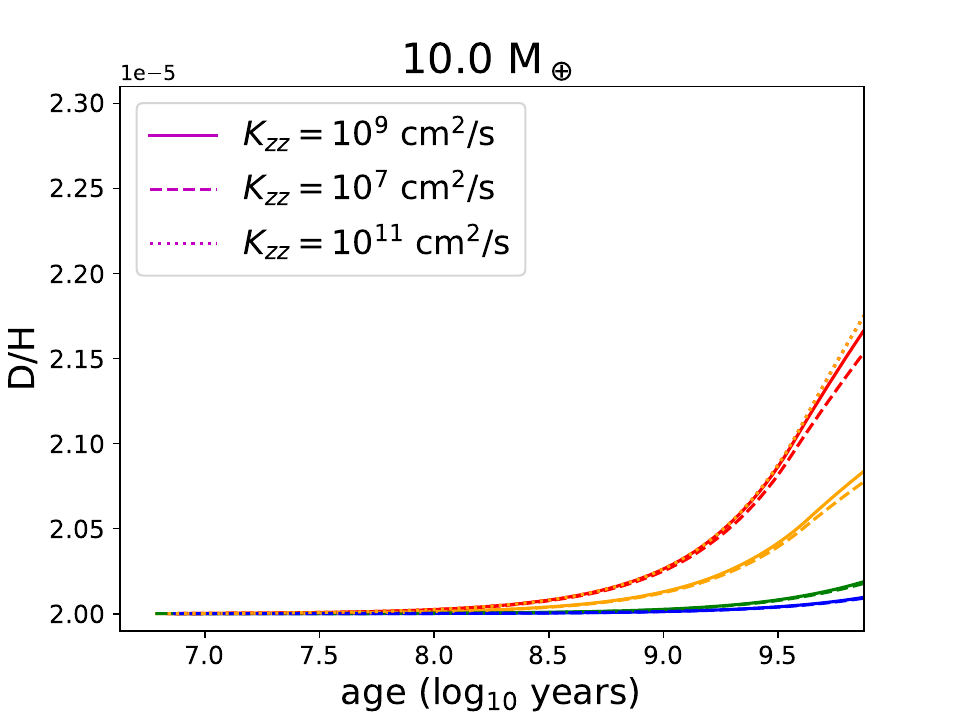} &
\includegraphics[width=0.35\linewidth]{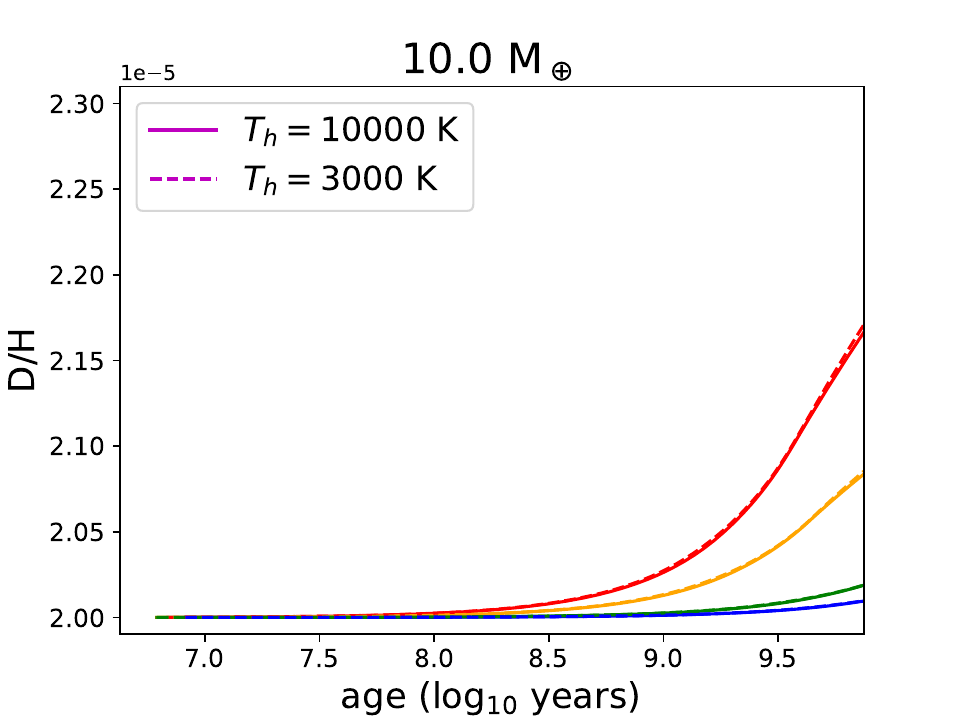} 
\end{tabular}
\caption{Dependence of D/H evolution on the free parameters  $\eta$, $K_{zz}$, and $T_h$. The solid curves illustrate the typical evolution with the fiducial parameters shown in Figure~\ref{fig:evol_0.2}, while the dashed and dotted curves present those with other values of free parameters. The curve for $\eta=0.75$ in the case of $f_{env}=0.005$ is not plotted in the upper left panel due to the unphysical radius anomaly in the simulation. The evolution is insensitive to the free parameters except in extreme cases for the high photoevaporation efficiency $\eta=0.75$ shown in the lower left panel, where the D/H value of the massive sub-Neptune with an initial thin atmosphere (e.g., $f_{env}=0.005$ and $M_p=10M_\oplus$) can be more enhanced at $t=7.5$ Gyr due to the faster diffusive escapes of both H and D during the late stage, when $t \gtrsim$1-2 Gyr.}
\label{fig:freeparam}
\end{figure}




\end{document}